\documentclass[sigconf]{acmart}
\usepackage{graphicx}
\usepackage{multirow}
\usepackage[table,xcdraw]{xcolor}
\usepackage{xspace}
\usepackage{makecell}
\usepackage{amsmath}
\usepackage{balance}
\usepackage{multirow}
\usepackage[normalem]{ulem}
\useunder{\uline}{\ul}{}
\usepackage{enumitem}
\usepackage{amsthm,amsmath}
\usepackage{mathrsfs}
\usepackage{booktabs}
\usepackage{booktabs} % 用于美观的表格线
\usepackage{multirow} % 用于多行合并
\usepackage{amsmath}  % 用于数学符号，如 \textbf 和 \underline
\usepackage{graphicx} % 如果需要插入图形可以用
\usepackage{algorithm}
\usepackage{algorithmic}
\usepackage{subfigure}
\usepackage{tcolorbox}
\usepackage{graphicx}
\usepackage{textcomp}

\newcommand{\icon}[2][1.1ex]{\raisebox{-.2\height}{\includegraphics[height=#1]{#2}}}

\usepackage{pifont}
\AtBeginDocument{%
  }

\setcopyright{acmlicensed}
\copyrightyear{2018}
\acmYear{2018}
\acmDOI{XXXXXXX.XXXXXXX}
\acmConference[Conference acronym 'XX]{Make sure to enter the correct
  conference title from your rights confirmation email}{June 03--05,
  2018}{Woodstock, NY}
\acmISBN{978-1-4503-XXXX-X/2018/06}

\begin{document}

\title{From Manipulation to Mistrust: Explaining Diverse Micro-Video Misinformation for Robust Debunking in the Wild}

\author{Zhi Zeng}
\authornote{Equal Contribution}
\affiliation{%
  \institution{School of Computer Science and Technology, MOEKLINNS Lab, Xi'an Jiaotong University}
  \city{Xi'an, Shaanxi}
  \country{China}}
\email{zhizeng@stu.xjtu.edu.cn}

\author{Yifei Yang}
\authornotemark[1]
\affiliation{%
  \institution{School of Computer Science and Technology, MOEKLINNS Lab, Xi'an Jiaotong University}
  \city{Xi'an, Shaanxi}
  \country{China}}
\email{yangyf001@stu.xjtu.edu.cn}

\author{Jiaying Wu}
\authornote{Corresponding authors.}
\affiliation{%
  \institution{National University of Singapore}
  \city{}
  \country{Singapore}}  
\email{jiayingw@nus.edu.sg}

\author{Xulang Zhang}
\affiliation{%
  \institution{Nanyang Technological University}
  \city{}
  \country{Singapore}}  
\email{xulang.zhang@ntu.edu.sg}

\author{Xiangzheng Kong }
\affiliation{%
  \institution{Xi'an Jiaotong University}
  \city{Xi'an, Shaanxi}
  \country{China}}
\email{kxz1582366422@stu.xjtu.edu.cn }
% \email{mazihan880@stu.xjtu.edu.cn }

\author{Herun Wan}
\affiliation{%
  \institution{ Xi'an Jiaotong University}
  \city{Xi'an, Shaanxi}
  \country{China}}
\email{wanherun@stu.xjtu.edu.cn}

\author{ Zihan Ma}
\affiliation{%
  \institution{ Xi'an Jiaotong University}
  \city{Xi'an, Shaanxi}
  \country{China}}
\email{mazihan880@stu.xjtu.edu.cn}

\author{Minnan Luo}
\authornotemark[2]
\affiliation{%
  \institution{ Xi'an Jiaotong University}
  \city{Xi'an, Shaanxi}
  \country{China}}
\email{minnluo@xjtu.edu.cn}
% \authornote{}

\renewcommand{\shortauthors}{Zhi Zeng et al.}


\begin{abstract}
The rise of micro-videos has reshaped how misinformation spreads, amplifying its speed, reach, and impact on public trust. Existing benchmarks typically focus on a single deception type, overlooking the diversity of real-world cases that involve multimodal manipulation, AI-generated content, cognitive bias, and out-of-context reuse. Meanwhile, most detection models lack fine-grained attribution, limiting interpretability and practical utility. To address these gaps, we introduce \textbf{WildFakeBench}, a large-scale benchmark of over 10,000 real-world micro-videos covering diverse misinformation types and sources, each annotated with expert-defined attribution labels. Building on this foundation, we develop \textbf{FakeAgent}, a Delphi-inspired multi-agent reasoning framework that integrates multimodal understanding with external evidence for attribution-grounded analysis. FakeAgent jointly analyzes content and retrieved evidence to identify manipulation, recognize cognitive and AI-generated patterns, and detect out-of-context misinformation. Extensive experiments show that FakeAgent consistently outperforms existing MLLMs across all misinformation types, while WildFakeBench provides a realistic and challenging testbed for advancing explainable micro-video misinformation detection.\footnote{Data and code are available at: \url{https://github.com/Aiyistan/FakeAgent}.}
\end{abstract}





\maketitle

\begin{figure*}
    \centering{\includegraphics[width=\linewidth, height=10.3 cm, keepaspectratio]{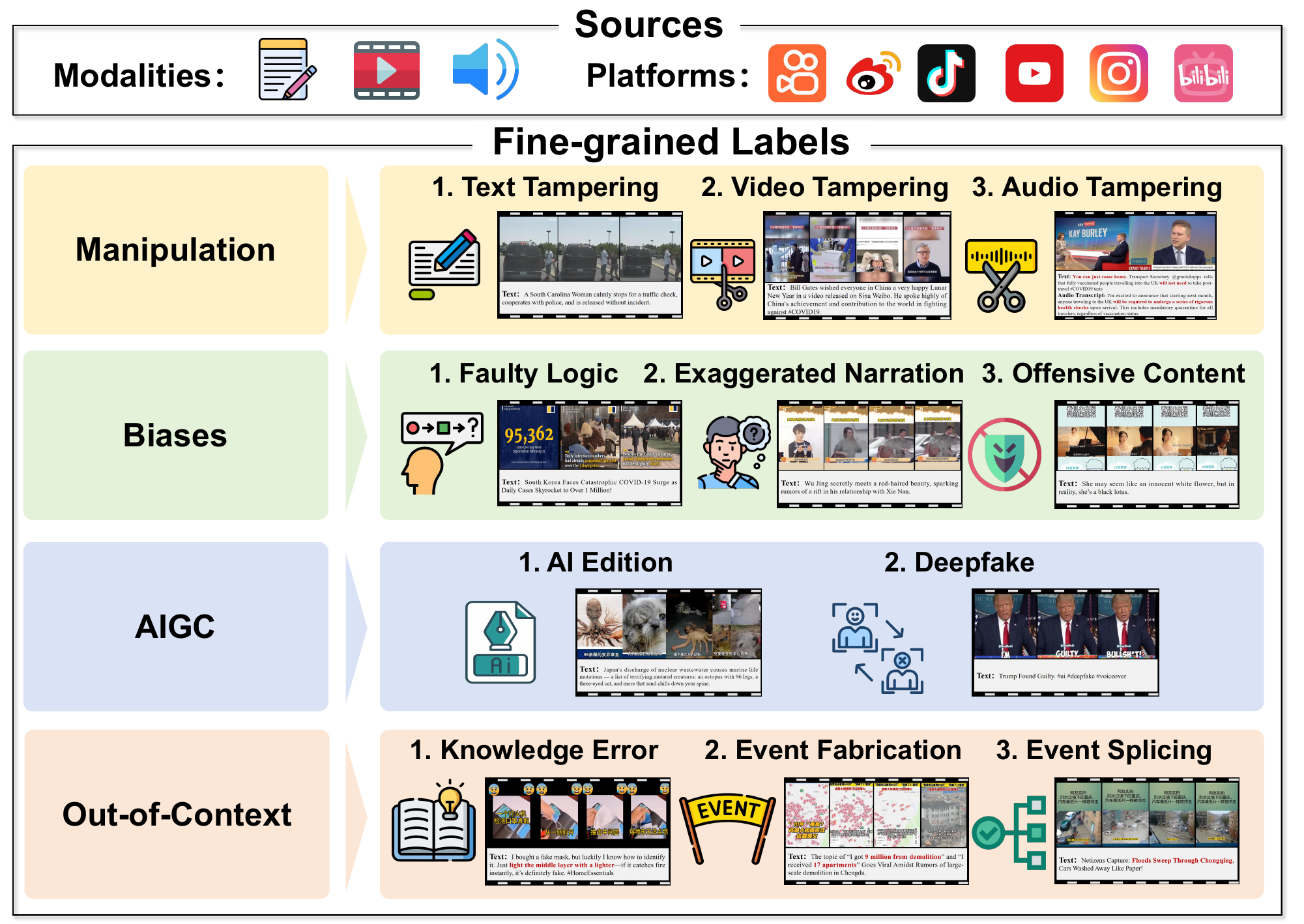}}
    \caption{WildFakeBench at a glance: over 10,000 real-world micro-videos capturing diverse forms of misinformation.}\label{intro}
\end{figure*}

\section{Introduction}

Micro-videos have redefined how misinformation spreads online, becoming a major medium for news consumption and public discourse \citep{walker2021news}. While these platforms enable open participation, they also accelerate the circulation of deceptive content, posing new challenges to public trust in the information ecosystem. Compared with text-based misinformation \citep{kuangye2024bayesian}, deceptive micro-videos involve \textbf{diverse and intertwined forms of deception}, including multimodal manipulation \citep{fmnv,COVID-VTS}, AI-generated content \citep{socialdf,MMaigc,mmfakebench}, cognitive bias exploitation \citep{fakeve,multihateclip}, and out-of-context reuse of authentic footage \citep{mdam3,amg,newsclippings}. Advanced creators exploit these variations to mislead viewers without leaving obvious inconsistencies.

Despite steady progress in multimodal misinformation detection \citep{FANVN,TikTec,fakesv,fakett,imol,acl-fake-news-video,mmdfnd,lu2025dammfnd,tong2025dapt}, two major gaps remain.
First, existing \textbf{benchmarks} are confined to specific deception types, such as AIGC or visual manipulation, and therefore fail to capture the complex and hybrid nature of real-world misinformation.
Second, emerging \textbf{reasoning-based methods}, particularly those using multimodal large language models (MLLMs) \citep{www-fnvd,www25fnvd,Fact-Checking-fnvd,mdam3}, can generate natural-language explanations but often hallucinate and lack verifiable attribution to external evidence, limiting their reliability for fact-checkers and practitioners.

To bridge these data and reasoning gaps, we present two complementary contributions that together advance reliable, evidence-based detection of real-world micro-video misinformation.
We first introduce \textbf{WildFakeBench}, a large-scale benchmark of over 10,000 real-world micro-videos that capture diverse and intertwined deceptive strategies. It provides a unified testbed for analyzing both perceptual and reasoning aspects of misinformation across manipulation, cognitive bias, AI generation, and contextual distortion.
On top of this resource, we further develop \textbf{FakeAgent}, a multi-agent reasoning framework that detects and explains misinformation through attribution-grounded analysis. By jointly examining multimodal content and external evidence, it generates transparent reasoning chains that enhance both interpretability and reliability. Together, these contributions establish a unified foundation for studying and mitigating misinformation in the wild.

As illustrated in Figure~\ref{intro}, WildFakeBench organizes deceptive strategies into four categories: (1) \textbf{Manipulation}, involving altered visual, textual, or audio elements that distort perception \citep{fmnv}; (2) \textbf{Cognitive Biases}, exploiting logical or psychological cues that mislead interpretation \citep{fakeve,multihateclip, kuangye, wan2025truth, zeng1, zeng2}; (3) \textbf{AIGC}, synthetic or edited content generated by AI tools to amplify influence or fear \citep{socialdf,Ukraine_conflict}; and (4) \textbf{Out-of-Context}, authentic footage misrepresented through spliced or misleading narratives \citep{zengMM}. It aggregates content from six major social platforms and provides expert-annotated, fine-grained veracity categories, supporting systematic evaluation of both perceptual and reasoning-based detection models.

Building on this foundation, FakeAgent uses a \textbf{Delphi-inspired multi-agent design} \citep{delphiagent} to produce transparent and evidence-grounded reasoning chains. Instead of opaque binary predictions, FakeAgent jointly analyzes multimodal content and open-world knowledge to (1) detect manipulations across modalities, (2) distinguish cognitive-bias and AI-generated semantics, and (3) retrieve and attribute supporting evidence for out-of-context misinformation. By integrating multi-view knowledge reasoning with explicit attribution, FakeAgent improves both the accuracy and interpretability of misinformation detection, paving the way for more transparent and reliable multimodal reasoning frameworks.

\begin{figure}
    \centering{\includegraphics[width=\linewidth, height=7.9 cm, keepaspectratio]{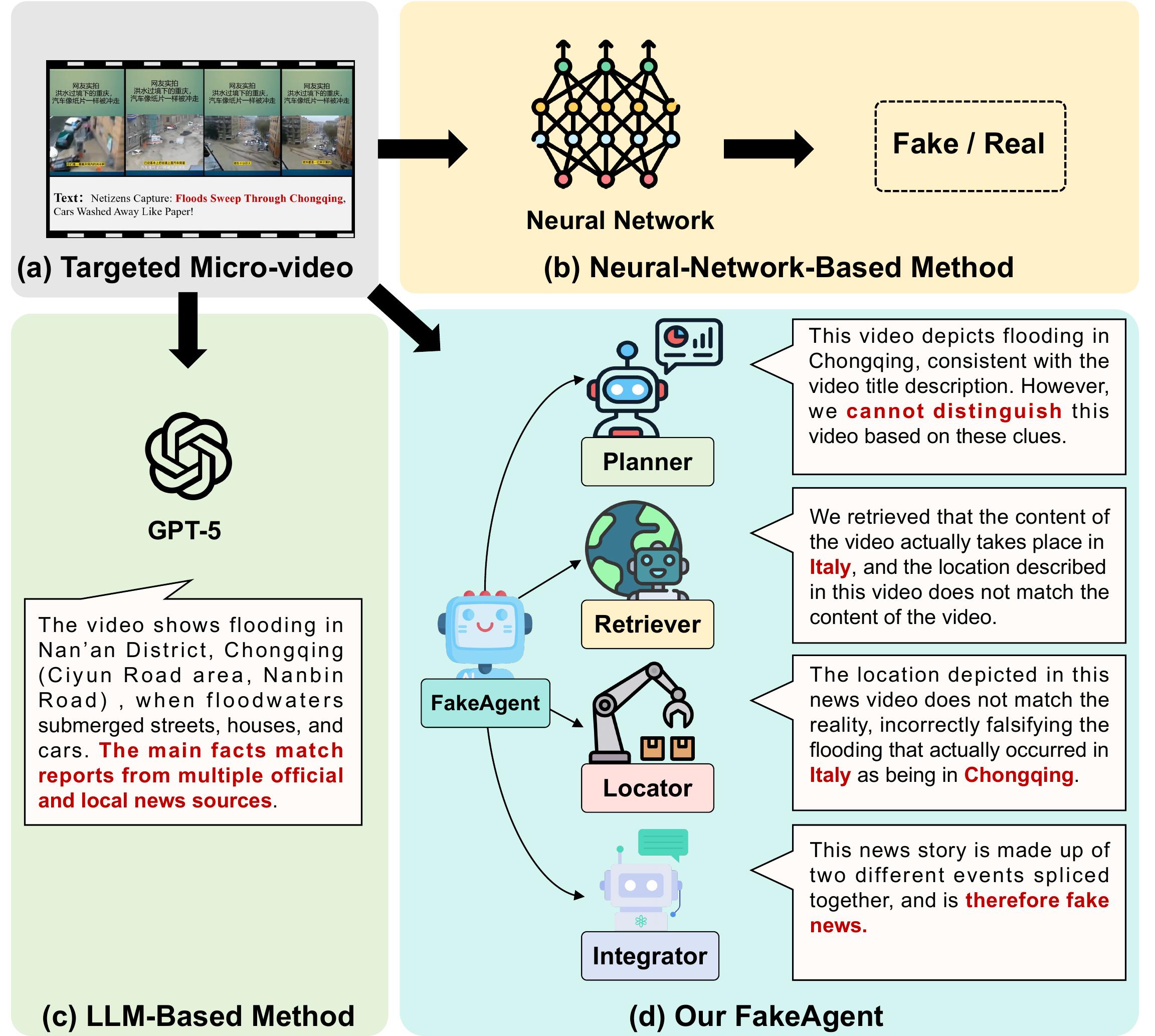}}
    \caption{Comparison of our proposed FakeAgent, neural-network-based and LLM-based methods.  }\label{intro-method}
\end{figure}

\section{Related Work}
% \jy{in case of space limit issues, consider shortening this section. your related work has taken too much space.}
\subsection{Benchmarks}
With the advancement of social media \citep{lixiang2025dual, lixiang2025scalegnn, lixiang2025umgad, ma2025bridging}, several benchmarks have been proposed to advance research in micro-video misinformation detection, as summarized in Table~\ref{tab:fake_news_datasets}. FVC \citep{FVC} was the first large-scale micro-video misinformation dataset, collecting textual titles and videos from different platforms like YouTube and Twitter. Several researchers \citep{palod2019misleading, hou2019towards} extended this effort by extracting misinformation micro-videos from Facebook and Twitter. Additionally, Serrano et al. \citep{COVID-19_misinformation} and Shang et al. \citep{TikTec} focused on Covid-19, creating English-language datasets on TikTok. Different from these single-domain datasets, Bu et al. \citep{fakett} designed FakeTT, a TikTok dataset spanning multiple domains, enhancing misinformation detection across contexts. To address the lack of non-English resources, Qi et al. \citep{fakesv} built the largest Chinese short misinformation micro-video dataset, incorporating multimodal information to support multimodal misinformation detection.

With the advancements in MLLMs, recent benchmarks \citep{mdam3, socialdf} introduce AI-generated or AI-editing content to reflect the diversity of real-world situations. However, these datasets cannot fully capture the diversity of misinformation in the wild, overlooking real-world out-of-context misinformation \citep{newsclippings, amg} that extends beyond the original knowledge boundaries of humans or detection models, which may lead to irreparable consequences.

To construct a comprehensive benchmark capturing the diversity of misinformation in the wild, we propose WildFakeBench, a multi-source misinformation micro-video attribution benchmark, WildFakeBench, which enables more comprehensive and challenging evaluation.

\begin{table*}
    \centering
    
    \setlength{\tabcolsep}{6pt}
    \renewcommand{\arraystretch}{1.2}
    \caption{
    Comparison of micro-video misinformation benchmarks. \textbf{WildFakeBench} spans the longest period, includes the most videos, and covers the most diverse deception types and sources. Source platforms: YT (YouTube), TW (Twitter), FB (Facebook), TT (TikTok), BB (Bilibili), WB (Weibo), DY (Douyin), IS (Instagram), KS (Kuaishou).
    }
    \label{tab:fake_news_datasets}
    \vspace{-0.3cm}
 
    % 7 columns: Datasets | Time Span | Misinformation/Real | #Video | Type | In Wild | Source
    \begin{tabular}{lcccccc}
     \Xhline{1.5pt}
     % \rule{0pt}{3ex} 
       {Datasets} & {Time Span} & \thead{\#Post\\(Misinformation/Real)} & {\#Video} & 
        {Type} & {In Wild} & {Source}
        \\ \hline
        % PolitiFact \citep{GossipCop} & --- & 135/224 & 0 & 1 & $\surd$ & PF  \\ 
        % GossipCop \citep{GossipCop}  & --- & 2,036/7,974 & 0 & 1 & $\surd$ & PF  \\ 
        FVC\citep{FVC}               & -2018 & 2,916/2,090 & 5,006 & 1 & $\surd$ & YT/TW  \\ 
        (Palod et al. 2019)\cite{palod2019misleading} & 2013-2016 & 123/423 & 546 & 1 & $\surd$ & FB  \\ 
        (Hou et al. 2019)\cite{hou2019towards} & -2019 & 118/132 & 250 & 1 & $\surd$ & TT  \\ 
        (Serrano et al. 2020)\citep{COVID-19_misinformation} & -2020 & 113/67 & 180 & 1 & $\surd$ & YT  \\ 
        (Choi and Ko 2021)\citep{FANVN} & -2021 & 902/903 & 1,805 & 1 & $\surd$ & YT  \\ 
        (Shang et al. 2021)\citep{TikTec} & -2020 & 226/665 & 891 & 1 & $\surd$ & TT  \\ 
        (Li et al. 2022)\citep{li2022cnn} & 2014-2015 & 210/490 & 700 & 1 & $\surd$ & BB  \\ 
        FakeSV\citep{fakesv} & 2017-2022 & 1,827/1,827 & 3,654 & 1 & $\surd$ & DY/KS  \\ 
        FakeTT\citep{fakett} & 2019-2024 & 1,172/819 & 1,991 & 1 & $\surd$ & TT  \\ 
        MMFakeBench\citep{mmfakebench} & -& 3,300/7,700 & 0 & 3 & $\times$ & Synthetic   \\ 
        MDAM$^3$\citep{mdam3} & - & 90,000/0 & 90,000 & 4 & $\times$ & Synthetic  \\ 
        \hline
        WildFakeBench (Ours) & 2017-2025 & 4,122/5,985 & 10,107 & 10 & $\surd$ & \thead{WB/DY/KS\\YT/IS/BB}  \\ 
    \Xhline{1.5pt}
    \end{tabular}
    \vspace{-0.3cm}
\end{table*}

\subsection{Methods}
% \jy{start directly from deep learning approaches to save some space.}

\paragraph{\textbf{Micro-video Misinformation Detection.}} Early researchers \citep{COVID-19_misinformation, hou2019towards} initially used handcrafted features from video titles and comments to identify misinformation. As deep learning advances, several studies \citep{li2022cnn, FANVN, TikTec, acl-fake-news-video, coling_short_video}, used neural network methods for automatic feature extraction. While multimodal approaches have further enriched this field, SV-FEND \citep{fakesv}, a Transformer-based model, was proposed to integrate multimodal knowledge for misinformation detection. Similarly, TwtrDetective \citep{COVID-VTS} incorporated cross-media consistency. Moreover, NEED \citep{need} employed graph attention networks to incorporate event-related and debunking knowledge, enhancing contextual awareness. FakingRecipe \citep{fakett} explored material preferences and editing processes to identify distinctive misinformation patterns. Additionally, Zeng et al. \citep{zengMM} proposed multimodal multi-view debiasing framework for mitigating bias in micro-video misinformation identification.

\paragraph{\textbf{MLLM-based Multimodal Misinformation Detection.}} MLLM-based misinformation detection task typically aims to incorporate MLLM's world knowledge into the analysis of multimodal misinformation. Early researchers applied MLLMs to identify multimodal misinformation, such as EARAM \citep{www_EARAM}, MMDIR \citep{MMDIR}, Sniffer \citep{sniffer}, and FKA-Owl \citep{Fka-owl}. Several studies \citep{DiFaR, mmfakebench, sigir_CAMERED, cikm-fnd} incorporate external multi-view knowledge into enhancing misinformation detection by offering additional knowledge insights through role-based responses. Although these approaches offer some reasoning capabilities, they are prone to ``hallucinations'', leading to insufficient authenticity and reliability of the explanations. To address this, MLLMs are designed as enhancers via Chain-of-Thought (Cot) \citep{www-fnvd}, Retrieval-Augmented Generation (RAG) \citep{naacl-RAG-misinfor}, reinforcement learning \citep{Fact-R1}and external evidence \citep{Fact-Checking-fnvd, mdam3} to enhance reliability.

These approaches primarily focus on multimodal information integration and MLLM-based Knowledge enhanced misinformation detection while overlooking the autonomous ability to explore and integrate information in the wild. To address these, we propose the FakeAgent approach that a multi-agent framework that integrates cross-modal knowledge with the autonomous exploration and integration of real-world external evidence for more reliable and comprehensive detection.

\section{WildFakeBench Curation}
\label{sec:dataset}
We introduce \textbf{WildFakeBench}, the first large-scale benchmark designed to support explainable micro-video misinformation detection across diverse social platforms.\footnote{The ethical statement for data collection and annotation is provided in Appendix~\ref{Ethical_statement}.}

% In this section, we present WildFakeBench, the first benchmark designed to support explainable micro-video misinformation detection on various platforms.\footnote{ Ethical statement for data collection and annotation is provided in Appendix~\ref{Ethical_statement}.}

\subsection{Data Collection and Filtering}

To ensure the credibility of annotations and consistency with verified sources, we curated micro-videos referencing fact-checked events from \textit{PolitiFact} and the \textit{China Internet Joint Rumor-Refuting Platform}, two nationally recognized authorities in misinformation verification. The dataset encompasses six major platforms: Weibo, Douyin, Kuaishou, YouTube, Instagram, and Bilibili, covering a period from 2017 to 2025.

We retained only micro-videos containing verifiable claims to ensure relevance and factual grounding. To reduce redundancy and prevent risks of data leakage, rigorous textual similarity filtering was applied to eliminate near-duplicate content while preserving topic diversity.

% To ensure the credibility of the annotated micro-videos and their alignment with authoritative debunking sources, we constructed a dataset enriched with additional event data from PolitiFact and the China Internet Joint Rumor-Refuting Platform, both nationally recognized authorities on misinformation verification. The micro-videos were collected from six widely used platforms, namely Weibo, Douyin, Kuaishou, YouTube, Instagram, and Bilibili, covering the period from 2017 to 2025.

% To maintain relevance and avoid redundancy, we retained only micro-videos containing verifiable claims. Rigorous textual similarity filtering was then applied to eliminate near-duplicate articles, thereby enhancing content diversity and minimizing the risk of potential data leakage.

\begin{figure}
    \centering{\includegraphics[width=\linewidth, height=8.1 cm, keepaspectratio]{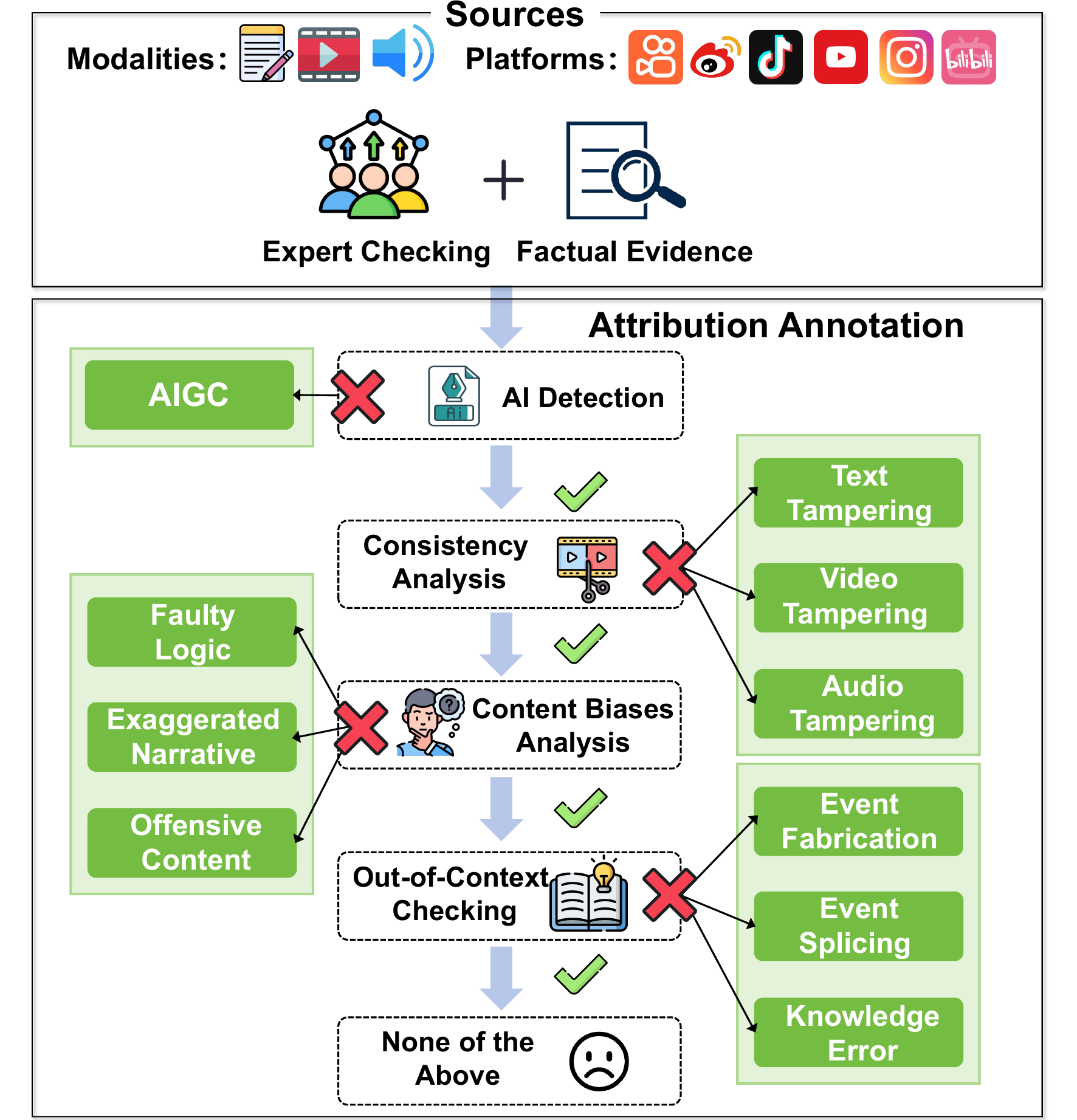}}
    \caption{Overview of the data annotation process.}\label{data_annotation}
\end{figure}

\begin{figure}[t]
	\centering
	\subfigure[The distribution of real vs. misinformation micro-videos.]{
		\label{cl1} %% label for second subfigure
		\includegraphics[width=\linewidth, height=3.5 cm, keepaspectratio]{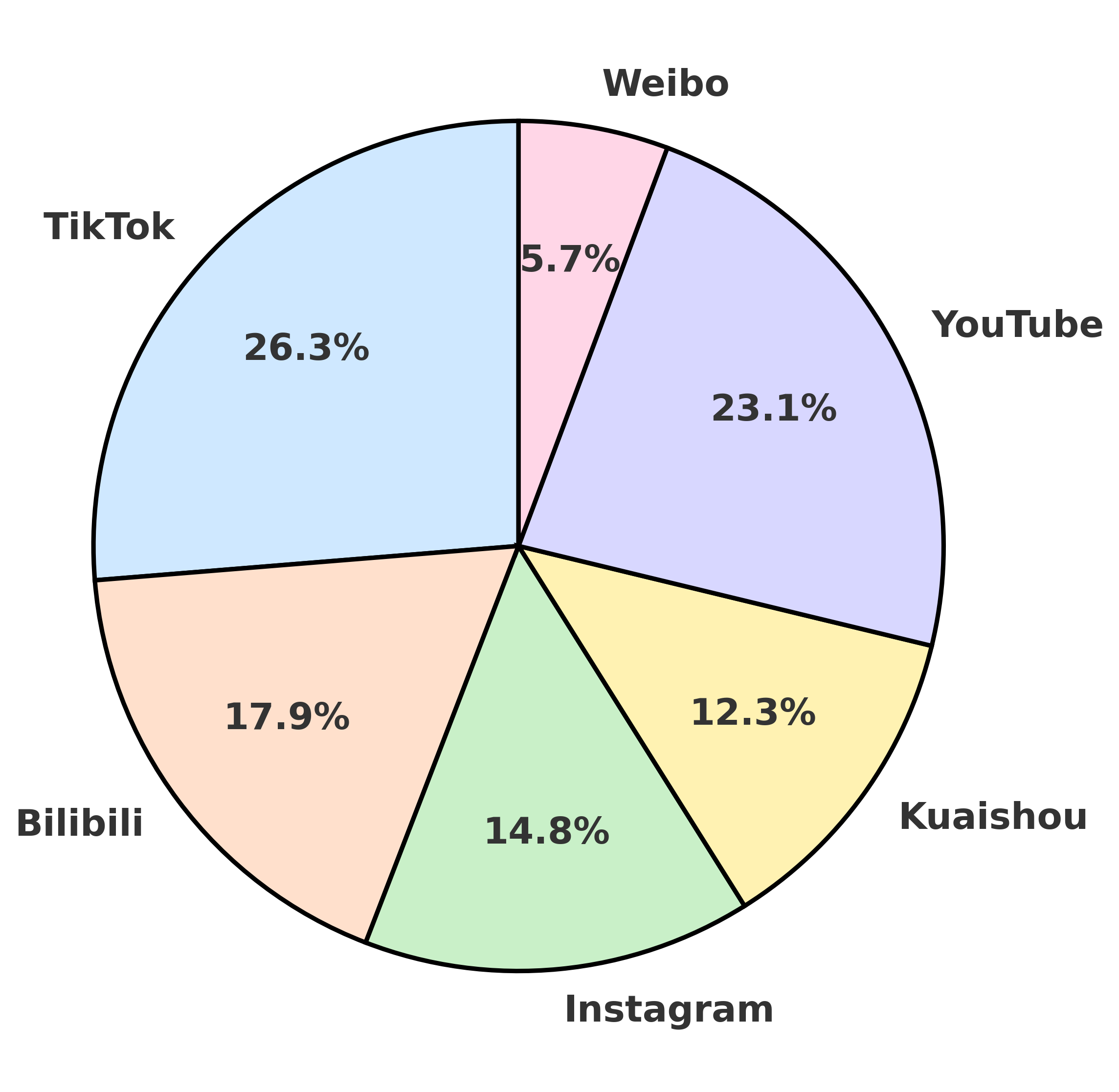}}
	\hspace{0in}
	\subfigure[The type percentage of the misinformation micro-videos.]{
		\label{cl2} %% label for second subfigure
		\includegraphics[width=\linewidth, height=2.8 cm, keepaspectratio]{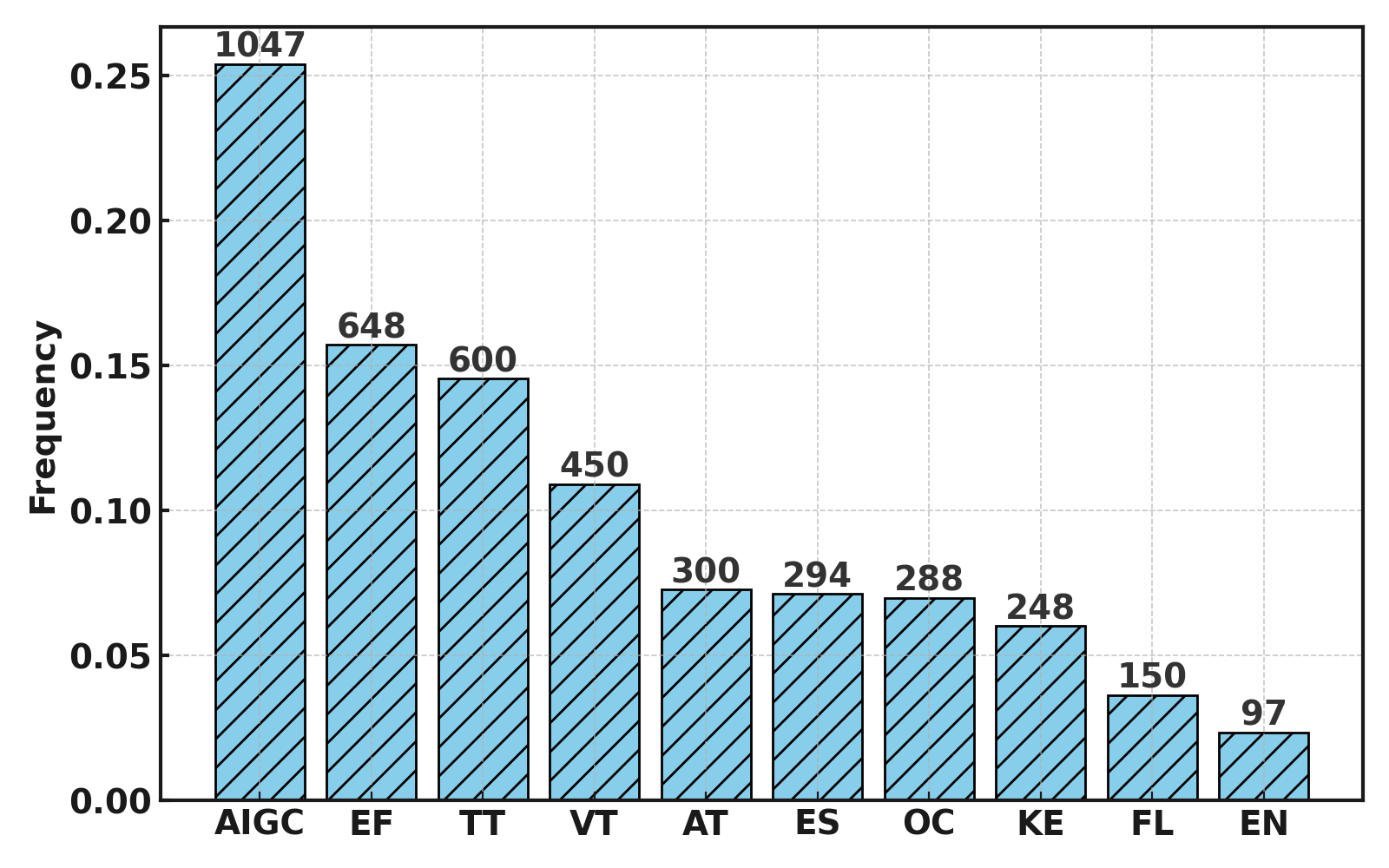}}
  \setlength{\abovecaptionskip}{0.1cm} %调整图片标题与图距离
\setlength{\belowcaptionskip}{-0.1cm} %调整图片标题与下文距离
\caption{Data analysis of our WildFakeBench.}\label{distribution}
\end{figure}

\begin{figure}[t]
	\centering
	\subfigure[Misinformation]{
		\label{cl1} %% label for second subfigure
		\includegraphics[width=\linewidth, height=2.6 cm, keepaspectratio]{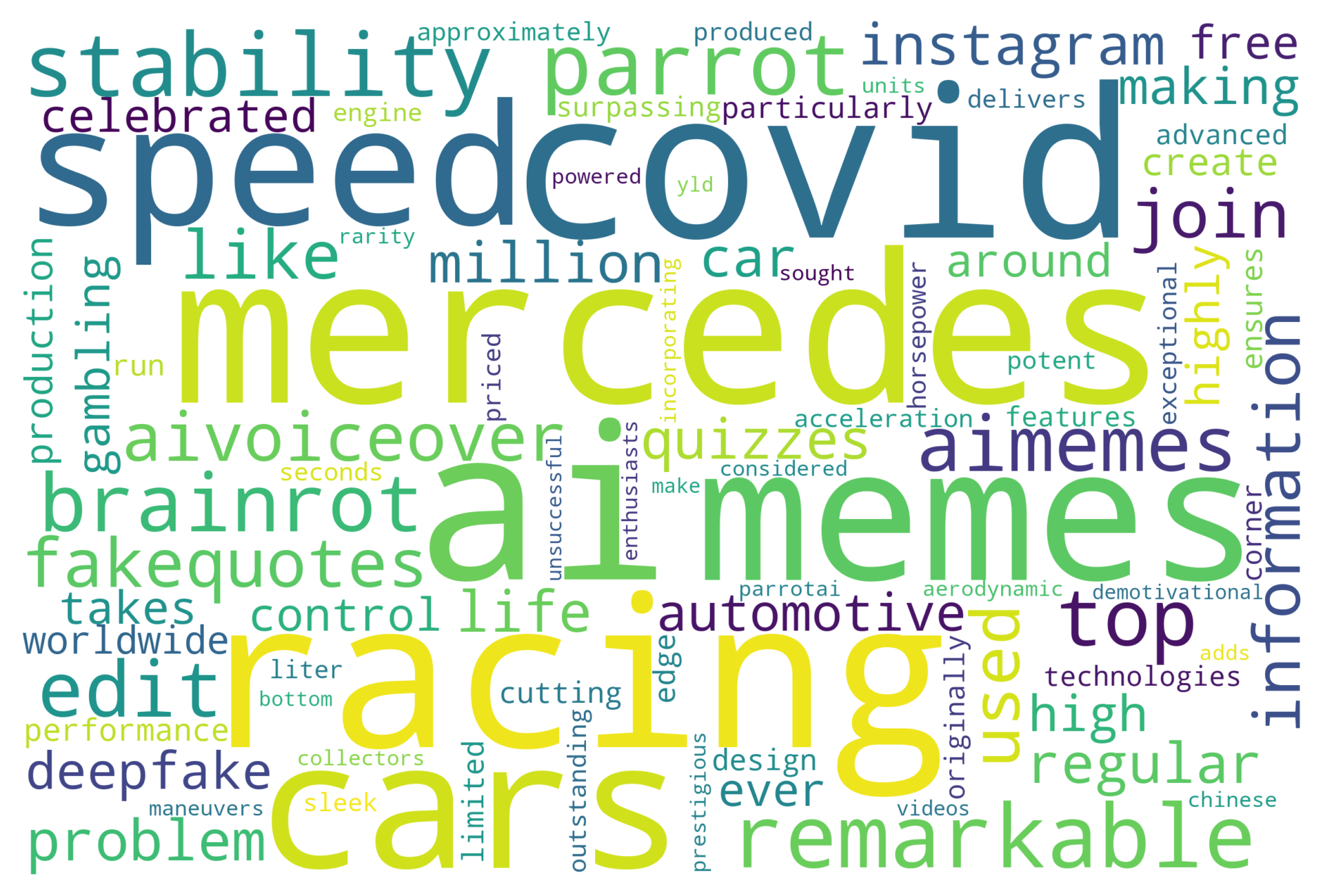}}
	\hspace{0in}
	\subfigure[Real]{
		\label{cl2} %% label for second subfigure
		\includegraphics[width=\linewidth, height=2.6 cm, keepaspectratio]{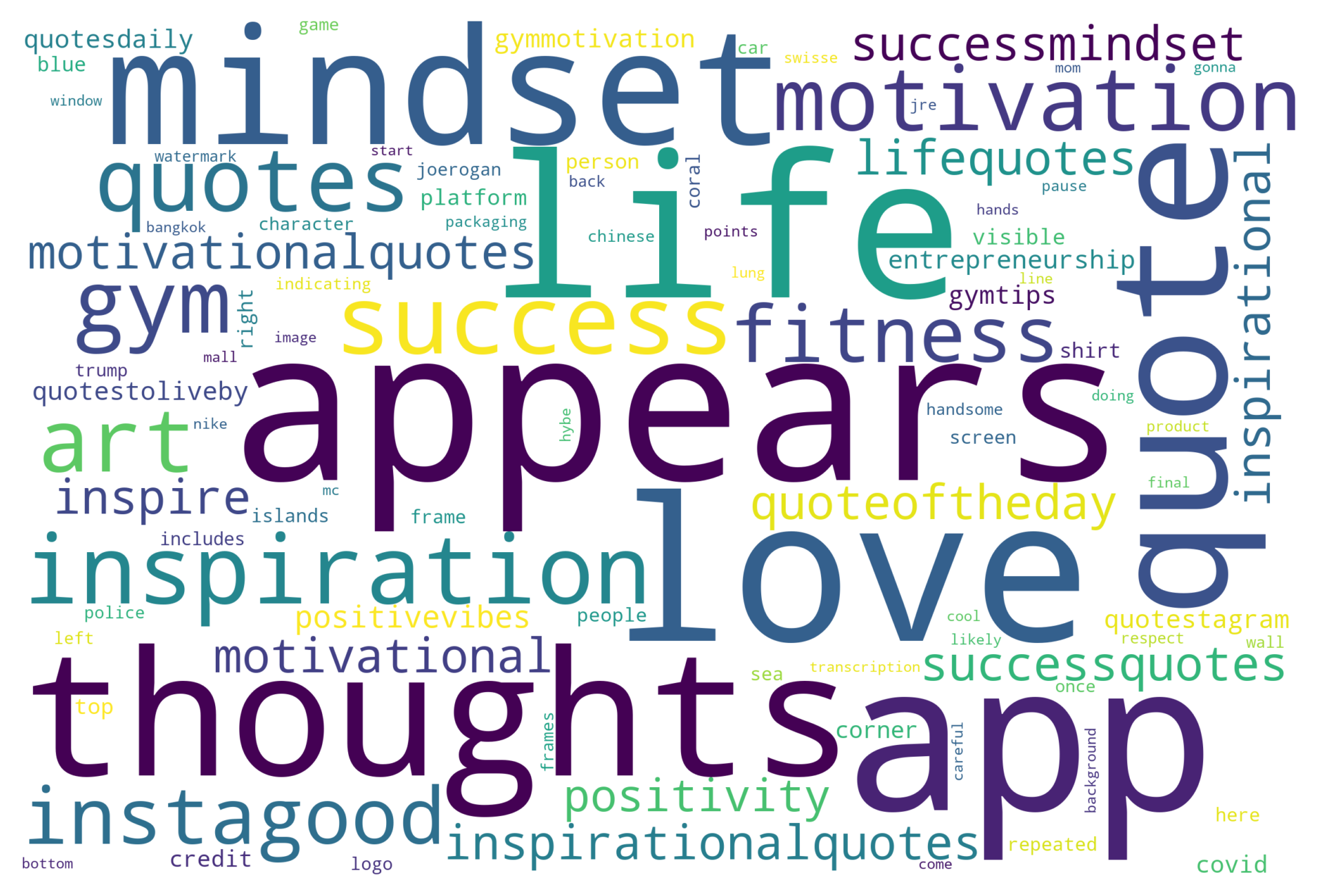}}
  \setlength{\abovecaptionskip}{0.1cm} %调整图片标题与图距离
\setlength{\belowcaptionskip}{-0.1cm} %调整图片标题与下文距离
\caption{Domain-specific word clouds in WildFakeBench.}\label{wordcloud}
\end{figure}

\begin{figure*}[t]
    \centering{\includegraphics[width=\linewidth, height=12 cm, keepaspectratio]{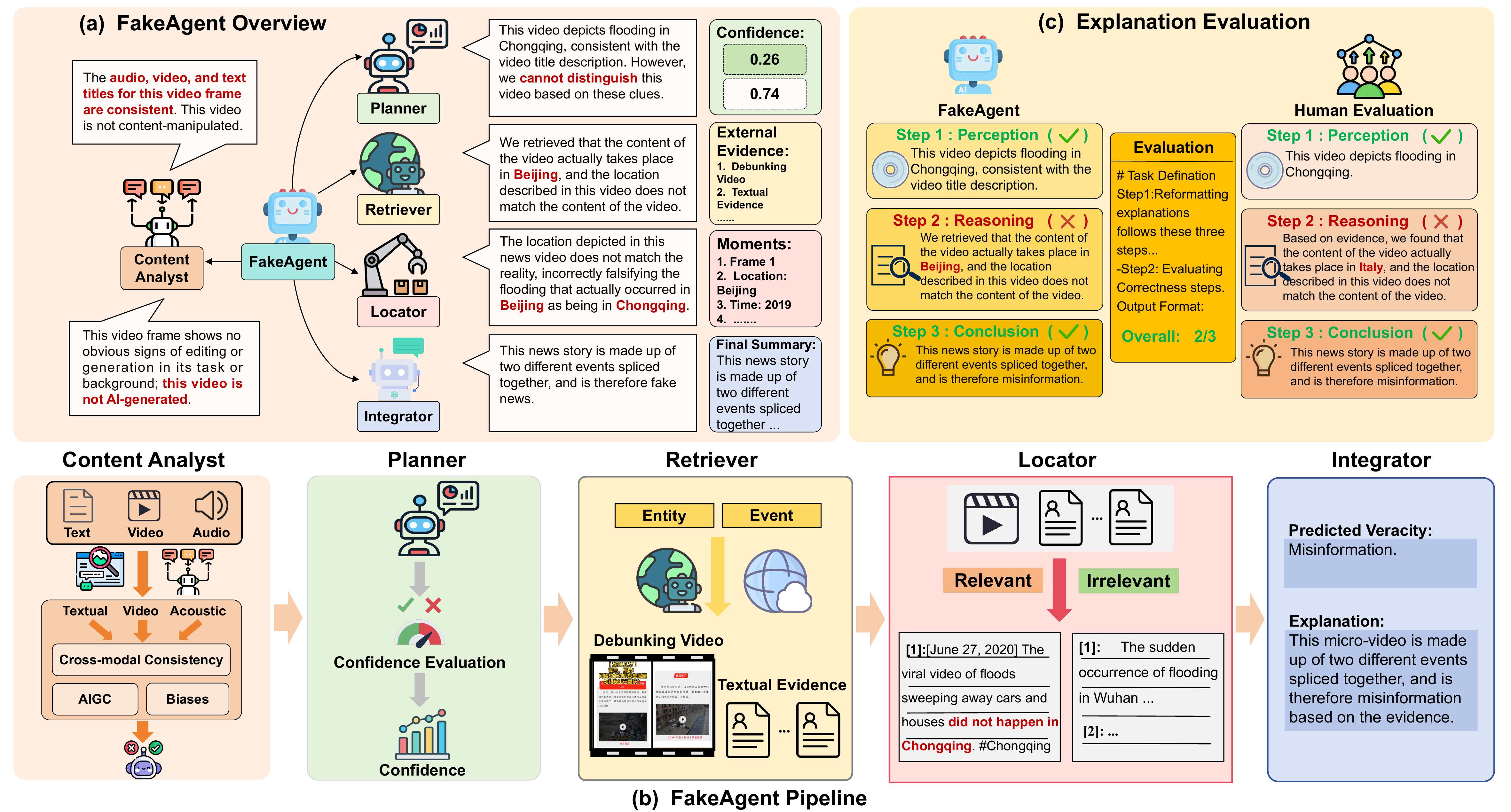}}
    \caption{Overview of our proposed FakeAgent framework. }\label{fakeagent}
\end{figure*}

\subsection{Data Annotation}

Unlike prior benchmarks that rely solely on binary veracity labels \citep{fakesv,fakett} or use synthetic content \citep{mdam3,mmfakebench}, WildFakeBench adopts a fine-grained, multi-dimensional annotation framework grounded in factual evidence (Figure~\ref{data_annotation}). In addition to binary Real/Fake labels, each sample receives a \textit{fine-grained attribution label} describing the mechanism of deception.

Our annotation follows a four-stage reasoning process, with 10 subtypes capturing distinct deceptive strategies:

\begin{itemize}[leftmargin=*]
    \item \textbf{Stage 1: AI-Generated Content (AIGC).} Identify whether the micro-video is (1) \textbf{AIGC}, such as content synthesized or heavily edited using generative AI tools to simulate real-world events or evoke emotional reactions.
    \item \textbf{Stage 2: Multimodal Manipulation.} Detect inconsistencies or falsifications across modalities: (2) \textbf{Text Tampering (TT)} modifies captions, titles, or on-screen text to misrepresent the visual or factual content. (3) \textbf{Video Tampering (VT)} alters or splices video segments to visually distort the original narrative. (4) \textbf{Audio Tampering (AT)} manipulates voiceovers, background sounds, or overlays to fabricate claims or emotional cues.
    \item \textbf{Stage 3: Cognitive Biases.} Capture psychological or rhetorical strategies used to influence perception: (5) \textbf{Faulty Logic (FL)} introduces misleading causal or correlational reasoning, such as false analogies or post hoc conclusions. (6) \textbf{Exaggerated Narration (EN)} employs overstated or sensational language to heighten engagement. (7) \textbf{Offensive Content (OC)} leverages implicit hate speech or personal attacks to evoke moral outrage.
    \item \textbf{Stage 4: Out-of-Context Manipulation.} Identify cases where authentic material or partial truths are used deceptively: (8) \textbf{Knowledge Error (KE)} misinterprets legitimate information, often framed with pseudo-scientific or misleading narratives. (9) \textbf{Event Fabrication (EF)} invents events without factual basis, often supported by fabricated visuals or commentary. (10) \textbf{Event Splicing (ES)} combines unrelated real-world clips or scenes to construct a false narrative.
\end{itemize}
Representative examples with corresponding debunking evidence are provided in Figure~\ref{intro}. Approximately 1.3\% of micro-videos that could not be confidently categorized were excluded. Each sample was independently annotated by at least three experts, and final labels were determined through unanimous consensus. The annotation experts included twelve individuals with academic or master’s degrees in computer science and social science.

\subsection{Data Analysis}
Figure~\ref{distribution} presents the distribution of misinformation sources and fine-grained attribution categories in WildFakeBench, highlighting its broad coverage across deception types, modalities, and platforms. The multi-level annotation framework and platform diversity enable comprehensive study of how misinformation manifests across global short-form video ecosystems. Additionally, different micro-video types exhibit distinct topical and linguistic characteristics. To further illustrate the linguistic patterns across micro-video categories, we generate word clouds depicting the most frequent vocabulary within each type (Figure~\ref{wordcloud}).

\section{Problem Definition}
\label{sec:problem_defn}

Given a micro-video dataset $\mathcal{D} = \{(x_i, y_i^d, y_i^a)\}_{i=1}^M$ with three modalities: text, video, and audio, each micro-video is represented as $x_i = (x_i^t, x_i^v, x_i^a)$. Each micro-video is assigned an attribution type label $y_i^a \in \{\text{type}_1, \dots, \text{type}_K\}$, where $K$ is the number of attribution types. Each micro-video is also assigned a veracity label $y_i^d \in \{0, 1\}$, where $y_i^d = 0$ indicates that the micro-video is real, and $y_i^d = 1$ indicates that it is misinformation.

\paragraph{\textbf{Task 1 (Multi-source micro-video misinformation detection).}}
Given $\mathcal{D} = \{(x_i, y_i^d, y_i^a)\}_{i=1}^M$, the task of multi-source micro-video misinformation detection aims to identify whether a micro-video $x_i$ is misinformation ($y_i^d = 1$) or real ($y_i^d = 0$).

\paragraph{\textbf{Task 2 (Multi-source micro-video misinformation explanation).}}
Given $\mathcal{D} = \{(x_i, y_i^d, y_i^a)\}_{i=1}^M$, the task of multi-source micro-video misinformation explanation aims to generate an explanation $e_i$ for the detection of micro-video $x_i$, which evaluates and interprets the reasoning process leading to the veracity prediction.

\section{Methodology}
As illustrated in Figure~\ref{fakeagent}, {FakeAgent} simulates the collective intelligence of multiple reasoning agents that collaboratively perceive micro-video content, retrieve external evidence, and evaluate veracity.
By combining perception and reasoning, it jointly analyzes multimodal content to detect both direct manipulation and subtle cognitive bias.
The system further integrates adaptive evidence retrieval from authoritative sources with internal reasoning, producing interpretable and evidence-grounded misinformation detection.

% FakeAgent simulates the collective intelligence of multiple agents by perceiving micro-video content, retrieving relevant external evidence, and evaluating the veracity of micro-videos, as illustrated in Figure~\ref{fakeagent}. Specifically, FakeAgent addresses both content manipulation and cognitive biases by jointly analyzing content consistency and modeling underlying human cognitive biases. It then adaptively retrieves reliable knowledge from authoritative media sources and trusted online repositories, and synthesizes this evidence to accurately detect misinformation in micro-videos in an interpretable manner.

\subsection{Multimodal Content Understanding}

Large language models (LLMs) have demonstrated strong analytical capabilities for misinformation detection \citep{dell,wu2025seeing,cikm-GenFEND}.
However, micro-videos frequently employ multiple deceptive strategies, such as multimodal manipulation \citep{fakeve,fmnv,tifs-Ma25} and implicit semantic deception \citep{multihateclip,wwwhate}, which challenge purely text-based or single-modality detection.
To address these challenges, FakeAgent introduces a \textbf{content analysis agent} that applies Chain-of-Thought (CoT) reasoning \citep{cot} across three levels: (1) identifying AI-generated content (AIGC), (2) analyzing multimodal content consistency, and (3) modeling deeper cognitive biases. This design enables the generation of diverse veracity-related rationales $\mathcal{R}=\{R_j\}_{j=1}^N$.

Each rationale $R_j$ is produced through structured multi-turn interactions. The MLLM first analyzes the micro-video from a designated perspective, such as ``evaluate the consistency between text and visuals'', yielding an intermediate rationale $R_j$. The content analysis agent then synthesizes these rationales into an internal veracity conclusion $\mathcal{C}$.

To encourage reasoning diversity, we design three representative prompt templates. At the AIGC level, prompts guide the model to identify possible AI-generated artifacts \citep{MMaigc}. At the content consistency level, prompts direct the model to assess alignment among textual, visual, and acoustic modalities. At the cognitive bias level, prompts elicit reasoning about logical coherence and offensive framing following prior studies \citep{fakeve}. Detailed templates are included in Appendix~\ref{Prompts}.
These representative prompts highlight FakeAgent’s adaptability and reasoning diversity and can be easily extended to other domains.

\subsection{External Evidence Reasoning}

While MLLMs exhibit strong internal reasoning abilities, they often struggle to identify out-of-context misinformation without access to external evidence \citep{mdam3, zeng2025understand}. To address this, {FakeAgent} introduces an adaptive evidence reasoning module that dynamically determines when and how to retrieve external information.

The process begins with a \textbf{planner agent} that estimates confidence in its internal reasoning based on the rationale set $\mathcal{R}$ \citep{Self-DC}. If the confidence score is insufficient for reliable prediction, the planner activates a \textbf{retriever agent} to acquire external evidence.

Given the title or key text of a micro-video $x_i^t$, the retriever constructs a query and gathers supporting information from authoritative media sources and verified repositories such as Wikipedia. The retrieved evidence corpus is defined as: (Eq.1)

% Although multimodal large language models (MLLMs) demonstrate strong capabilities in identifying misinformation in micro-videos, they often fail to distinguish out-of-context misinformation without access to external evidence \citep{mdam3}. To address this limitation, we introduce an adaptive confidence evaluation together with an external retrieval mechanism for adaptive evidence acquisition. 

% Specifically, FakeAgent employs a planner agent to determine whether external evidence is required by adaptively estimating its confidence \citep{Self-DC} based on the veracity-related rationale set $\mathcal{R}$.  

% In the absence of sufficient confidence to make a decision, FakeAgent employs a retriever agent to enable real-time access to external knowledge. The retriever agent integrates reliable and authoritative sources (e.g., Wikipedia and other trusted repositories) for fact-checking. Given the title of a micro-video $x_i^t$, the retriever agent constructs a query and retrieves an evidence corpus:
\begin{equation}
\mathcal{E} = \operatorname{retriever}(x_i^t, S_k),
\end{equation}
where $\mathcal{E}$ represents the retrieved evidence and $S_k$ denotes the top-$K$ ranked results from trusted domains.

To ensure semantic alignment between evidence and micro-video content, a \textbf{locator agent} further filters and localizes the most relevant information:
\begin{equation}
\mathcal{F} = \operatorname{locator}(x_i^t, \mathcal{E} ).
\end{equation}

Here, $\mathcal{F}$ contains the filtered and context-aligned evidence segments that directly support veracity assessment.

This multi-agent collaboration allows FakeAgent to adaptively integrate internal reasoning with external validation. The detailed prompt templates for planner, retriever, and locator agents are provided in Appendix~\ref{Prompts}.

% Here, $\mathcal{F}$ contains the filtering and detailed evidence and the detailed prompt templates of planner, retriever and locator agents are presented in the Appendix~\ref{Prompts}.

\subsection{Multi-view Evidence Integration}
Given the internal conclusion $\mathcal{C}$ and the filtered external evidence $\mathcal{F}$, FakeAgent integrates both into a unified evidence set:
% Based on the internal veracity-related conclusion $\mathcal{C}$ and filtered external evidence $\mathcal{F}$, we integrate it into a unified evidence set:
\begin{equation}
\mathcal{E}_{agg} = \{\mathcal{C}, \mathcal{F}\},
\end{equation}
where $\mathcal{E}_{agg}$ combines internal reasoning with retrieved evidence, providing complementary views for misinformation detection across modalities.
An \textbf{integrator agent} then synthesizes this aggregated evidence to produce the final decision:
% where $\mathcal{E}_{agg}$ combines content-internal knowledge with external evidence, thus providing multi-view clues for detecting misinformation in micro-videos across multiple modalities. Finally, an integrator agent synthesizes the aggregated evidence to derive a conclusion regarding the micro-video:
\begin{equation}
\mathcal{C} = \operatorname{integrator}(\mathcal{E}_{agg}),
\end{equation}
where $\mathcal{C}=\{\hat{y},e\}$ includes the predicted veracity $\hat{y}$ and its corresponding explanation $e$.

\begin{table*}[t]
  \centering
  \renewcommand\tabcolsep{5.6pt}
  \renewcommand\arraystretch{1.1}
  \caption{\textbf{Main results (\%).} Performance of all models on the primary evaluation metric, Micro-Acc. Misinformation types are abbreviated: TT (Text Tampering), VT (Video Tampering), AT (Audio Tampering), FL (Faulty Logic), EN (Exaggerated Narration), OC (Offensive Content), KE (Knowledge Error), EF (Event Fabrication), ES (Event Splicing), and AIGC (AI-Generated Content).}
  \label{Table-category}
  \scalebox{0.9}{
    % l  +  ccc(Modality) + ccc(Basic) + ccc(Sentiment) + ccc(New) + c(Fine) + c(Mean)
    \begin{tabular}{lccc|ccc|ccc|ccc|c|c}
     \Xhline{1.5pt} 
      & \multicolumn{3}{c|}{\textbf{Modality}}
      & \multicolumn{3}{c|}{\textbf{Content Manipulation}}
      & \multicolumn{3}{c|}{\textbf{Semantic Biases}}
      & \multicolumn{3}{c|}{\textbf{Out-of-Context}} % <-- 新增三列分组（可改名）
      & {\textbf{AIGC}} & \multirow{2}{*}{\textbf{Mean}} \\
      & A & V & T
      & TT & VT & AT
      & FL & EN & OC
      & KE & EF & ES % <-- 新增列名（可改）
      & AIGC & \\
      \hline
      \rowcolor{gray!20}
      \multicolumn{15}{c}{\textbf{Smaller-Parameter MLLMs}} \\
      % OneLLM        & $\surd$ & $\times$ & $\surd$
      %               & 25.52 & 17.21 & 28.32
      %               & 64.01 & 54.09 & 63.39
      %               & -- & -- & --
      %               & 22.25 & 41.14 \\
      % SECap         & $\surd$ & $\times$ & $\surd$
      %               & 40.95 & 52.46 & 25.56
      %               & 55.76 & 54.18 & 59.51
      %               & -- & -- & --
      %               & 36.97 & 46.64 \\
      % PandaGPT      & $\surd$ & $\times$ & $\surd$
      %               & 33.57 & 39.04 & 31.91
      %               & 66.06 & 61.33 & 62.93
      %               & -- & -- & --
      %               & 31.33 & 46.84 \\
      Qwen2-Audio-7B (Direct)   & $\surd$ & $\times$ & $\surd$
                    & 41.02 & 39.20 & 34.33 
                    & 70.42 & 43.30 & 22.92 
                    & 35.48 & 40.28 & 43.88 
                    & 21.39 & 49.29
                    \\
      % Qwen2-Audio-7B (CoT)        & $\times$ & $\surd$ & $\surd$
      %               & - & - & -
      %               & - & - & -
      %               & - & - & -
      %               & - & - \\
      % \textbf{FakeAgent} & $\surd$ & $\times$ & $\surd$
      %               & \textbf{72.94} & \textbf{73.41} & \textbf{56.63}
      %               & \textbf{83.46} & \textbf{80.74} & \textbf{82.99}
      %               & -- & -- & --
      %               & \textbf{59.98} & \textbf{72.18} \\
      % \hline
       InternVL-3-8B (Direct)& $\times$ & $\surd$ & $\surd$
                    & 14.83 & 1.78 & 6.67
                    & 29.33 & 28.87 & 24.65
                    & 24.19 & 19.14 & 11.22
                    &36.87 & 19.76 \\
      % InternVL3-14B & $\surd$ & $\surd$ & $\surd$
      %               & 67.50 & 16.00 & 31.33
      %               & 86.67 & 92.78 & 85.43
      %               & 90.32 & 81.64 & 73.13
      %               & 99.43 & 64.17 \\
      
       InternVL-2.5-8B (Direct)       & $\times$ & $\surd$ & $\surd$
                    & 14.33 & 12.00 & 13.33
                    & 13.33 & 65.94 & \textbf{81.94}
                    & 64.92 & 72.53 & \textbf{74.15}
                    &91.98 & 50.45 \\
        
        % InternVL-2.5-32f         & $\times$ & $\surd$ & $\surd$
        %             & 48.83 & 11.56 & 26.00
        %             & 56.00 & 60.82 & 72.92
        %             & 65.32 & 61.88 & 58.16
        %             &85.58 & 54.71 \\ 
     
      Qwen2.5-VL-7B (Direct) & $\times$ & $\surd$ & $\surd$
                    & 34.33 & 9.78  & 13.33
                    & 33.33 & 70.10 & 60.62
                    & 65.73 & 62.65 & 52.38
                    & \textbf{95.51} & 49.78 \\

     LLaVA-OneVision-7B (Direct) & $\times$ & $\surd$ & $\surd$
                    & 22.50 & 14.00 & 22.00
                    & 35.33 & 38.14 & 28.22 
                    & 25.00 & 25.62 & 30.27 
                    & 25.69  & 26.68 \\
      
      Qwen2.5-Omni-7B (Direct) & $\surd$ & $\surd$ & $\surd$
                    & 25.33 & 24.22 & 27.66
                    & 33.33 & 26.80 & 22.92
                    & 39.52 & 26.70 & 21.09
                    & 47.76 & 29.53 \\
      \hline
       \rowcolor{gray!20}
       \multicolumn{15}{c}{\textbf{Larger-Parameter MLLMs}} \\ 
         Gemma3-12B (Direct)    & $\times$ & $\surd$ & $\surd$
                    & 28.67 & 9.56 & 49.11
                    & 34.67 & 57.73 & 49.31
                    & 61.69 & 54.01 & 41.50
                    & 72.97 & 45.92\\
        InternVL-2.5-38B (Direct) & $\times$ & $\surd$ & $\surd$
                    & 48.83 & 11.56 & 15.33
                    & 56.00 & 60.82 & 72.92
                    & 65.32 & 61.88 & 58.16
                    &85.58 & 54.71 \\ 
     
      Qwen2.5-VL-32B (Direct) & $\times$ & $\surd$ & $\surd$
                    & 53.83 & 12.89 & 57.67
                    & 78.67 & 57.73 & 57.99
                    & 55.24 & 45.99 & 43.20
                    & 90.83 & 55.50 \\
         
      \hline
       \rowcolor{gray!20}
       \multicolumn{15}{c}{\textbf{MLLMs with VoT-based Prompt}} \\ 
     % InternVL-2.5-8B (VoT)        & $\times$ & $\surd$ & $\surd$
     %                & 23.67 & 7.78 & 10.33
     %                & 31.33 & 56.70 & 54.17
     %                & 50.81 & 47.53 &46.60
     %                & 61.60 & - \\
     % Qwen2.5-VL-7B (VoT)        & $\times$ & $\surd$ & $\surd$
     %                & 23.67 & 7.78 & 10.33
     %                & 31.33 & 56.70 & 54.17
     %                & 50.81 & 47.53 &46.60
     %                & 61.60 & - \\
    VideoLLaMA2-7B (VoT) & $\surd$ & $\surd$ & $\surd$
                    & 29.50 & 20.89 & 23.33
                    & 43.33 & 58.76 & 53.47 
                    & 53.56 & 50.31 & 45.58 
                    & 45.85  & 42.46 \\
      InternVL-2.5-38B (VoT)   & $\times$ & $\surd$ & $\surd$
                    & 66.67 & 17.33 & 29.67
                    & 72.00 & 72.23 & {81.25}
                    & 67.74 & 60.49 & 50.68
                    & 81.85 & 59.99\\
     % InternVL-2.5-38B (VoT) & $\times$ & $\surd$ & $\surd$
     %                & 46.83 & 12.67 & 20.67
     %                & 48.00 & 65.98 & 68.75
     %                & 66.53 & 49.38 & 37.41
     %                &98.38 & 51.46 \\ 
     
      Qwen2.5-VL-32B (VoT) & $\times$ & $\surd$ & $\surd$
                    & 49.33 & 45.45 & 25.67
                    & 69.33 & 73.20 & 59.72
                    & 68.55 & 57.25 & 48.64
                    & 55.87 & 55.30 \\
    \hline
       \rowcolor{gray!20}
       \multicolumn{15}{c}{\textbf{Closed-source MLLM}} \\ 
       GPT-4o (Direct)  & $\times$ & $\surd$ & $\surd$
                    & 58.50 & 16.22 & 32.67
                    & \textbf{79.33} & \textbf{84.54} & 69.44
                    & 77.02 & \textbf{72.53} & 66.67
                    & 94.46 & 65.14 \\
     \hline
       \rowcolor{gray!20}
       \multicolumn{15}{c}{\textbf{Our Proposed Approach}} \\                
     \textbf{FakeAgent-7B} & $\surd$ & $\surd$ & $\surd$
                    & \textbf{67.67} & \textbf{45.78} & \textbf{81.00}
                    & {70.00} & {68.04} & {63.19}
                    & \textbf{81.05} & {61.57} & {54.76}
                    & {93.22} & \textbf{68.63} \\
     
      \Xhline{1.5pt}
    \end{tabular}
  }
\end{table*}

\begin{table}
    % \centering
    
    \setlength{\tabcolsep}{6pt}
    \renewcommand{\arraystretch}{1.1}
      \captionof{table}{Ablation results (\%). Macro-level performance of different model variants on WildFakeBench.}
    \label{tab:ablation}
    \begin{tabular}{lcccc}
      \Xhline{1.5pt}
      Model & Acc & F1  & Macro-P & Macro-R \\
      \hline
      w/o Text & 61.34 & 54.74 & 58.16 & 55.86 \\
      w/o Video & 61.15 & 54.49 & 57.86 & 55.65 \\
      w/o Audio & 59.62 & 53.92 & 56.01 & 54.71 \\
      \hline
      w/o CKR & 60.49 & 52.29 & 56.36 & 54.16 \\
      w/o EER & 63.49 & 61.19 & 59.87 & 59.89 \\
      FakeAgent-7B   & \textbf{67.98} & \textbf{65.42} & \textbf{66.39} & \textbf{65.68} \\
      \Xhline{1.5pt}
    \end{tabular}
    \vspace{-0.3cm}
\end{table}

\section{Experiments}
% In this section, we conduct experiments to evaluate the effectiveness of our proposed model. Specifically, we aim to answer the following research questions:
In this section, we conduct extensive experiments to answer the following research questions:

% \jy{
\begin{itemize}[leftmargin=10 pt]
    \item \textbf{RQ1}~(\S\ref{performance}): Does FakeAgent improve micro-video misinformation detection?
    % \item \textbf{RQ2}~(\S\ref{performance}): Is WildFakeBnech is a challenge dataset?
    \item \textbf{RQ2}~(\S\ref{ablation}): How effective are FakeAgent's components?
    % \item \textbf{RQ3}~(\S Appendix \ref{complexity}): How does FakeAgent’s computational complexity compare to existing MLLMs?
    \item \textbf{RQ3}~(\S\ref{human}): Can FakeAgent generate high-quality explanation?
    \item \textbf{RQ4}~(\S\ref{casestudy}): What insights arise from FakeAgent’s case studies?
    % \item \textbf{RQ5}~(\S\ref{error}): What insights arise from FakeAgent’s error analysis?
\end{itemize}

\subsection{Experimental Settings}
\subsubsection{Baselines}
To evaluate both detection performance and explainability across diverse sources and misinformation types, we benchmark representative MLLMs. These include {InternVL-2.5} \citep{InternVL2.5}, {Qwen2.5-VL} \citep{qwen2.5vl}, {Qwen2-Audio} \citep{Qwen2-audio}, {Qwen2.5-Omni} \citep{Qwen2.5-omni}, {VideoLLaMA2} \citep{Videollama2}, {Gemma3} \citep{gemma-3}, {InternVL3} \citep{internvl3}, {LLaVA-OneVision} \citep{Llava-onevision}, and {GPT-4o} \citep{gpt4o}.
In addition, we implement enhanced inference variants based on the {Video-of-Thought (VoT)} paradigm \citep{vot}, which augments temporal reasoning for video-based misinformation detection.

% \footnote{The computational complexity analysis is provided in Appendix~\ref{complexity}.}.

% To test model's detection and explainable abilities on diverse sources and multi-type misinformation in the wild, we apply a series of MLLM models, such as InternVL-2.5 \citep{InternVL2.5}, Qwen2.5-VL \citep{qwen2.5vl}, Qwen2-audio \citep{Qwen2-audio}, Qwen2.5-Omni \citep{Qwen2.5-omni}, Gemma3 \citep{gemma-3}, InternVL3 \citep{internvl3}, LLavA-OneVision \citep{Llava-onevision} and GPT-4o \citep{gpt4}. Additionally, we design advanced inference MLLM models, leveraging Video-of-Thought (Vot) \citep{vot} for the MLLMs. 
\subsubsection{Evaluation Metrics}
In the era of MLLMs, micro-video misinformation detection requires assessing both classification accuracy and the reliability of model explanations. To further evaluate performance across different misinformation categories, we also report Micro-Accuracy for each deception type (Table~\ref{Table-category}).

\subsubsection{Implementation Details}
To enable the fair evaluation, we set the sampling hyperparameter of the off-the-shelf MLLMs, ``do\_sample = False'' or ``Temperature = 0'', to guarantee consistency in the prediction outputs. Additionally, we adopt the default setting of other hyperparameters such as ``max\_new\_tokens = 512''. For each micro-video, we uniformly sample 8 frames. For our FakeAgent, we utilize Qwen3 \citep{qwen3} as the LLM backbone, equipped with video and audio captioning capabilities \citep{internvl3, Qwen2-audio} to support comprehensive multimodal understanding. In the retriever agent, we set ``Top-K = 5''.
All experiments are conducted on four NVIDIA RTX 5880 Ada GPUs, each equipped with 48 GB of memory.
% \mathcal{S}^{\text{correct}} = \mathcal{S}^{\text{correct}}_{p} \cup \mathcal{S}^{\text{correct}}_{r}, 
% \quad
% \mathcal{S}^{\text{incorrect}} = \mathcal{S}^{\text{incorrect}}_{p} \cup \mathcal{S}^{\text{incorrect}}_{r}
% \end{equation}

\subsection{Main Results}\label{performance}

% \noindent\textbf{Macro Performance.} As shown in Table~\ref{tab:misinfo}, FakeAgent consistently outperforms all MLLM baselines, achieving a 6.16\% improvement in Macro-F1. Most MLLM baselines underperform on WildFakeBench, indicating that current models struggle to distinguish diverse and emerging misinformation types in real-world scenarios. These results demonstrate FakeAgent’s strong generalization and precise detection capabilities, while establishing WildFakeBench as a challenging benchmark for future research on micro-video misinformation detection.

% \textbf{Micro Performance.} Although existing MLLMs perform well in identifying AIGC, they face significant difficulty in detecting misinformation involving content manipulation, semantic bias, and out-of-context deception (Table~\ref{Table-category}). FakeAgent achieves higher accuracy across most categories, validating the effectiveness of its multi-agent collaboration in refining and integrating multimodal knowledge with external evidence.

\begin{itemize}[leftmargin=10 pt]
    % \item \textbf{Macro Performance.} As shown in Table~\ref{tab:misinfo}, FakeAgent consistently outperforms all MLLM baselines, achieving a 6.16\% improvement in Macro-F1. Most MLLM baselines underperform on WildFakeBench, indicating that current models struggle to distinguish diverse and emerging misinformation types in real-world scenarios. These results demonstrate FakeAgent’s strong generalization and precise detection capabilities, while establishing WildFakeBench as a challenging benchmark for future research on micro-video misinformation detection.
    \item \textbf{Micro Performance.} Although existing MLLMs perform well in identifying AIGC, they face significant difficulty in detecting misinformation involving content manipulation, semantic bias, and out-of-context deception (Table~\ref{Table-category}). With Video-of-Thought, MLLMs improve on most categories but show degraded performance on AIGC due to over-reasoning. FakeAgent achieves higher accuracy across most categories, validating the effectiveness of its multi-agent collaboration in refining and integrating multimodal knowledge with external evidence. 
     \item \textbf{Overall Conclusion.} Although our FakeAgent does not achieve the best performance in all subcategories, it attains the overall best results across all ten fine-grained categories, even surpassing GPT-4o. Moreover, FakeAgent contains fewer parameters than GPT-4o, Qwen2.5-VL-32B, and InternVL-2.5-38B, achieving an optimal balance between efficiency and performance. 
\end{itemize}
 
\begin{figure}
    \centering{\includegraphics[width=\linewidth, height=8 cm, keepaspectratio]{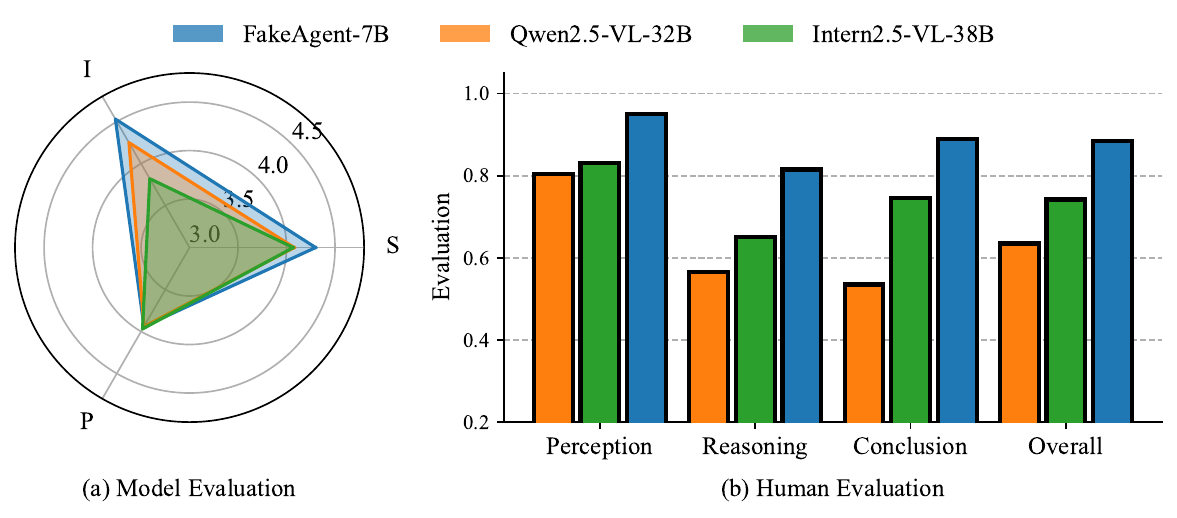}}
    \caption{Evaluation of model explanation quality using both automatic and human assessments.}\label{evaluation}
\end{figure}

\begin{figure*}[t]
	\centering
	\subfigure[Video Tampering Case]{
		\label{cs1} %% label for second subfigure
		\includegraphics[width=\linewidth, height=7.2 cm, keepaspectratio]{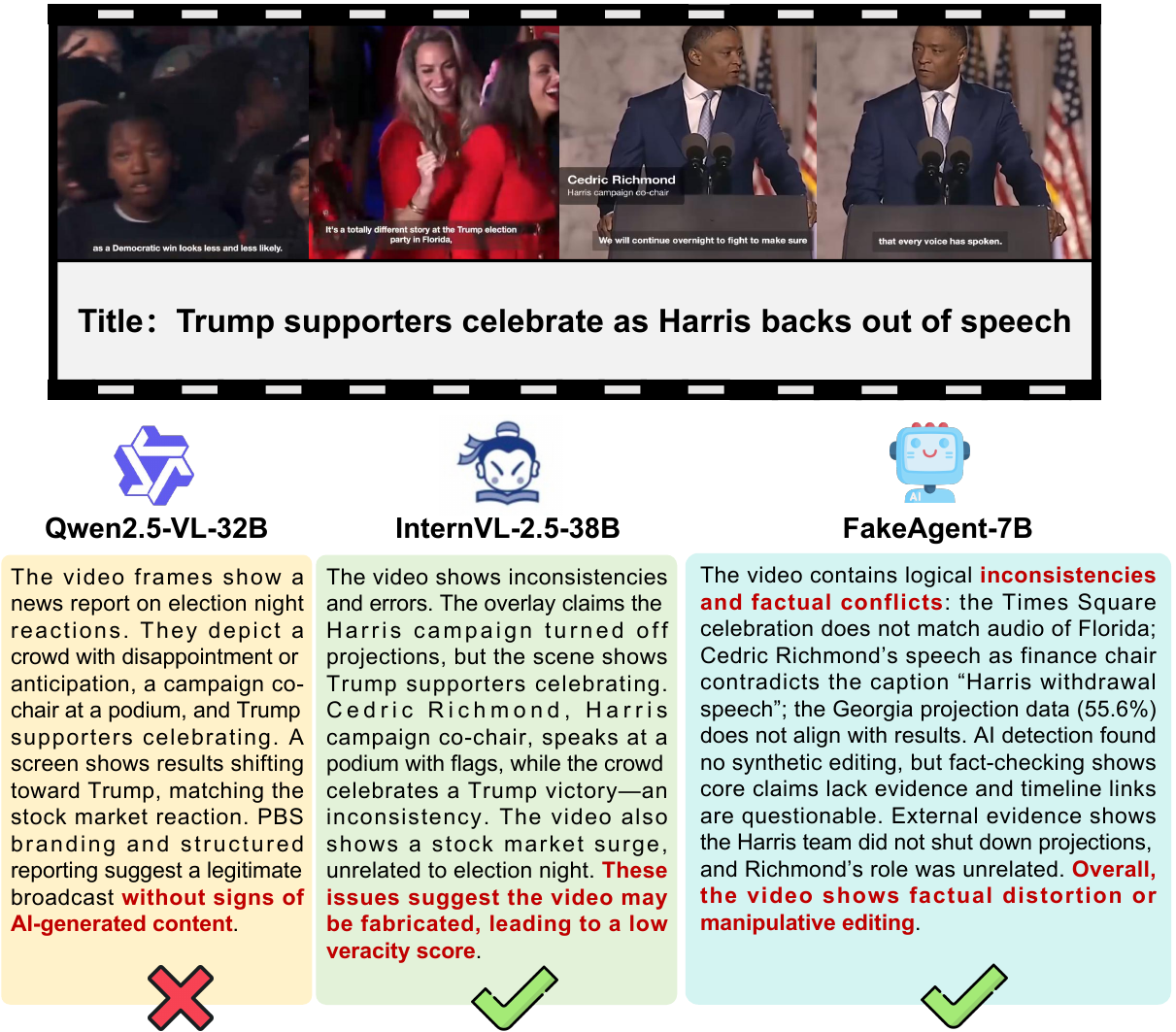}}
	\hspace{0in}
	\subfigure[Knowledge Error Case]{
		\label{cs2} %% label for second subfigure
		\includegraphics[width=\linewidth, height=7.2 cm, keepaspectratio]{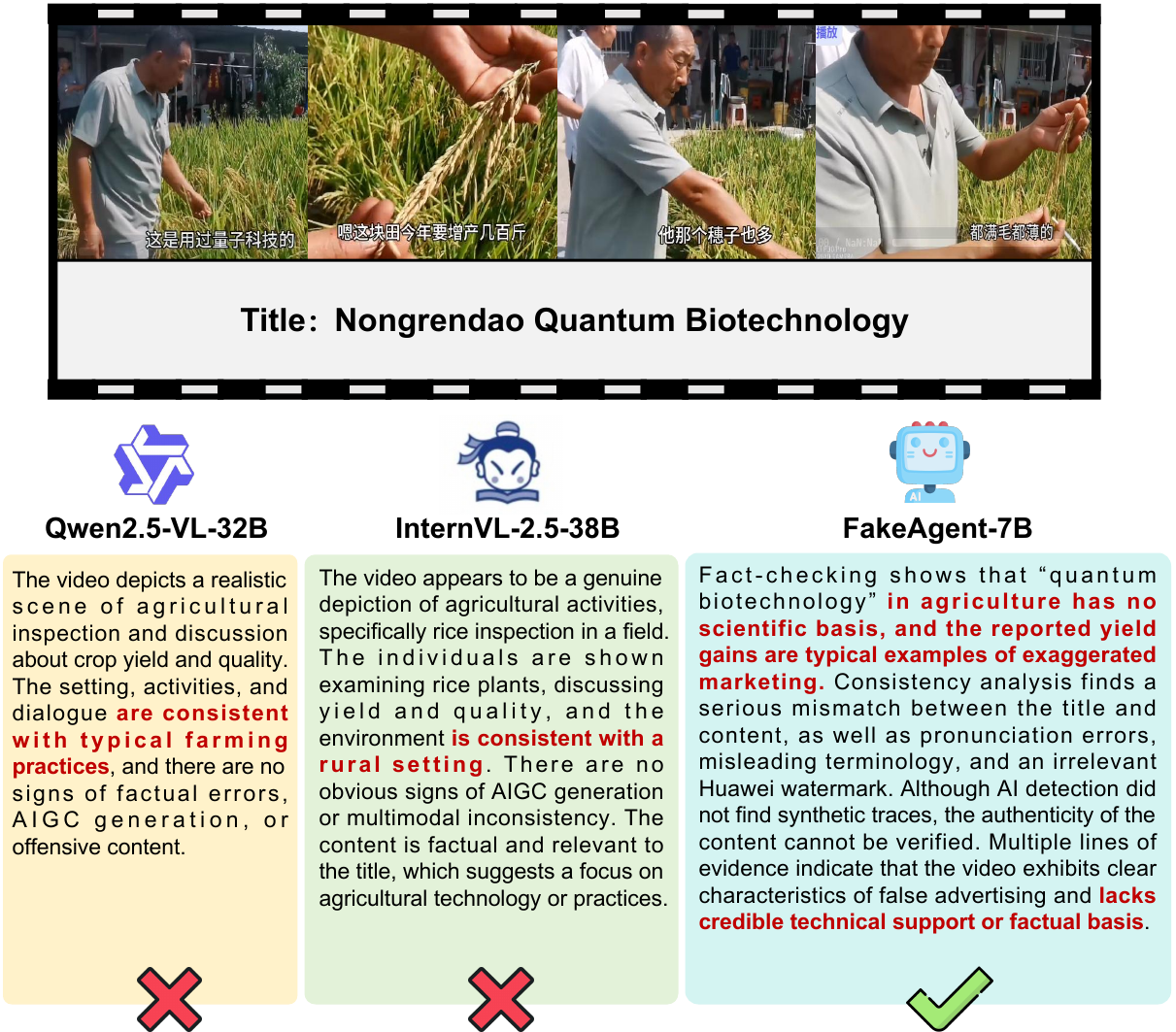}}
  \setlength{\abovecaptionskip}{0.1cm} %调整图片标题与图距离
\setlength{\belowcaptionskip}{-0.1cm} %调整图片标题与下文距离
\caption{Qualitative examples illustrating how FakeAgent detects and explains different types of micro-video misinformation.}\label{case_study}
\end{figure*}

\subsection{Ablation Study}\label{ablation}

To examine the contribution of different modalities, we construct seven internal variants of FakeAgent by removing textual, acoustic, or visual inputs.
The results in Table~\ref{tab:ablation} reveal that removing any modality leads to a clear performance drop, indicating that FakeAgent fully leverages multimodal information for effective micro-video misinformation detection.

We further evaluate the impact of each functional component through two ablation settings: \textbf{w/o CKR} (without cross-modal knowledge refining) and \textbf{w/o EER} (without external evidence retrieval).
As shown in Table~\ref{tab:ablation}, the complete {FakeAgent} consistently outperforms both variants, confirming that cross-modal knowledge refinement and external evidence retrieval jointly contribute to reliable detection and interpretable explanations.

\subsection{Explanation Quality Analysis}\label{human}

\subsubsection{Multi-view Evaluation}

We assess the explanatory quality of {Qwen2.5-VL-32B (VoT)}, {InternVL2.5-38B (VoT)}, and our {FakeAgent}.
Following prior work \citep{fmnv}, we adopt {G-Eval} \citep{G-Eval}, a reference-free, LLM-based evaluation framework that measures explanation quality across multiple dimensions.
Each explanation is rated by GPT-4o \citep{gpt4o} on three human-aligned dimensions: (1) \textbf{Persuasiveness (P)}, (2) \textbf{Informativeness (I)}, and (3) \textbf{Soundness (S)}, using a five-point Likert scale (1 = lowest, 5 = highest).

As shown in Figure~\ref{evaluation}(a), {FakeAgent} consistently surpasses larger MLLMs in both informativeness and soundness, confirming its ability to produce explanations that are more factual, detailed, and logically coherent.

% We evaluate the explanatory quality of outputs from Qwen2.5 VL-32B, InternVL 2.5-38B, and our FakeAgent. We adopt G-Eval \citep{G-Eval}, a reference-free evaluation based on large language models that can assess textual explanation quality from multiple perspectives. Following prior work, we measure explanatory quality of all samples \citep{fmnv} along three dimensions commonly used in human studies, Persuasiveness(P), Informativeness(I), and Soundness(S). Each dimension is scored on a five point Likert scale, where 1 denotes the worst and 5 denotes the best.

% Results in Figure~\ref{model_evaluation} demonstrate that FakeAgent consistenly outperform Larger MLLMs in informativeness, and soundness views, highlighting the FakeAgent outperforms larger MLLMs in explanation qulity.

\subsubsection{Human Evaluation}

To further examine fine-grained explanation quality, we conduct a human evaluation of Qwen2.5-VL-32B, InternVL2.5-38B, and FakeAgent across three perspectives:
\textbf{(1) Perception}, which measures the accuracy of describing video content;
\textbf{(2) Reasoning}, which evaluates the correctness of attribution and logical inference; and
\textbf{(3) Conclusion}, which assesses the accuracy of determining whether a micro-video constitutes misinformation.

We apply stratified sampling across the ten misinformation subcategories, selecting 20 short videos from each while ensuring diversity.
Each video is independently annotated by three human experts, and results are reported as the averaged scores.
As illustrated in Figure~\ref{evaluation}(b), {FakeAgent} consistently outperforms larger MLLMs across all three dimensions.
This shows the effectiveness of its retriever, locator, and integrator agents in refining knowledge and grounding explanations with external evidence.

% To evaluate the fine-grained quality of explanation in Qwen2.5-VL-32B, InternVL-2.5-38B and FakeAgent, we conduct multi-view human evaluation, including \textbf{1) Perception:} assessment of the method’s accuracy in describing video content, \textbf{2) Reasoning:} evaluating the accuracy of the method’s reasoning in the attribution process of identifying micro-videos, and \textbf{3) Conclusion:} evaluating the accuracy of the method in determining the conclusion of whether a micro-video is misinformation.
% For the ten subcategory of misinformation micro-videos, we use a stratified sampling method, 20 short videos were sampled from each category, while ensuring diversity among them. The experimental results are reported as the average of annotations provided by three human experts.

% Results in Figure~\ref{human_evaluation} demonstrate that FakeAgent consistenly outperform Larger MLLMs in perception, reasoning and conclusion views, highlighting the FakeAgent outperforms larger MLLMs in knowledge refining and external evidence retrieval. These findings support our design of retriever, locator and integrator agents as core components of FakeAgent.

\subsection{Case Study}\label{casestudy}
To qualitatively illustrate the perception and reasoning capabilities of {FakeAgent}, we analyze two cases of both multimodal manipulation and out-of-context misinformation.
Figure~\ref{cs1} shows that {FakeAgent} achieves finer-grained perception of multimodal falsification, accurately identifying textual and visual inconsistencies.
Figure~\ref{cs2} presents an out-of-context case where other MLLMs fail due to limited domain knowledge.
In contrast, FakeAgent autonomously retrieves relevant scientific evidence and constructs a coherent explanation, demonstrating the value of combining multimodal understanding with external knowledge retrieval.
These examples highlight the potential of WildFakeBench for advancing research on cross-modal reasoning and evidence-grounded explanation \footnote{The error analysis is provided in Appendix~\ref{error}.}.

% To intuitively demonstrate the perception and reasoning capabilities of FakeAgent, we conduct a qualitative analysis of both manipulation and out-of-context cases. Figure~\ref{cs1} illustrates the process of understanding multimodal content manipulation patterns, which shows our FakeAgent has the better fine-grained perception ability of multimodal falsification. Additionally, Figure~\ref{cs2} presents a out-of-context case where MLLMs cannot have the professional knowledge to detect and explain. In this case, FakeAgent autonomously identifies and retrieves relevant scientific evidence for effective explanation. This highlights the need for future studies on integrating cross-modal knowledge understanding and external knowledge retrieval and integration on WildFakeBench.

% While FakeAgent significantly improve the detection and explanation abilities of multi-type misinformation on various sources, it may lead to misjudgments regarding specialized and up-to-date academic knowledge, as shown in Figure~\ref{errorcase}. This requires enhancing the ability to detect misinformation that involves specialized academic knowledge. 

% In the future work, we aim to explore more adaptive and robust external knowledge exportation and integration mechanisms that enhance generalization across diverse type of misinformation. This enables our FakeAgent remains effective and accurate in evolving misinformation situations.

\section{Conclusion}
In this study, we highlight the importance of multi-source and multi-type approaches for detecting and explaining misinformation in real-world micro-videos.
To advance this goal, we introduce {WildFakeBench}, a large-scale benchmark featuring expert-annotated attributions across diverse deception forms, and {FakeAgent}, a multi-agent reasoning framework that integrates internal content understanding with external evidence for attribution-grounded analysis.
Extensive experiments show that {FakeAgent} achieves superior detection accuracy and delivers interpretable explanations, demonstrating strong generalization to previously unseen misinformation types. Together, these contributions provide a foundation for future research on evidence-grounded and explainable multimodal misinformation detection.

\begin{acks}
This work is supported by the Fundamental and Interdisciplinary Disciplines Breakthrough Plan of the Ministry of Education of China (No. JYB2025XDXM101), the National Natural Science Foundation of China (No. 62272374, No. 62192781), the Natural Science Foundation of Shaanxi Province (No.2024JC-JCQN-62), the State Key Laboratory of Communication Content Cognition under Grant No. A202502, the Key Research and Development Project in Shaanxi Province (No. 2023GXLH-024), and the Ministry of Education, Singapore, under its MOE AcRF TIER 3 Grant (MOE-MOET32022-0001). The China Scholarship Council also supports this research.
\end{acks}

\bibliographystyle{ACM-Reference-Format}
\bibliography{sample-base}

\appendix

\section{Legal and Ethical Statement}\label{Ethical_statement}

We strictly followed the data-use and scraping policies of all platforms involved in this study.
All annotators received formal training and were familiar with relevant data privacy and security regulations.
During annotation, only content related to public figures or public events was considered, and posts involving private individuals were excluded.

Our WildFakeBench dataset incorporates 2,393 video samples from FMNV \citep{fmnv}, and it is released under the Attribution NonCommercial ShareAlike 4.0 International license, CC BY NC SA 4.0. We will adopt this license to align with the licensing terms of several constituent datasets, thereby providing the same level of access.

To ensure privacy protection, all identifiable user information, including usernames and IDs, was anonymized.
We implemented safeguards throughout data processing and model training to prevent any leakage of personal data.
All collected data are securely stored on protected servers with access restricted to authorized research personnel only.

% We strictly adhered to the data scraping policies of each platform involved in this study. All annotators received thorough training and were familiar with relevant data privacy and security regulations. During the annotation process, only posts related to public figures or public events were selected, and content involving private individuals was excluded.

% To further protect privacy, all identifiable user information, including usernames and IDs, was anonymized. We implemented safeguards throughout data processing and model training to prevent any potential leakage of personal data. All collected data is securely stored on protected servers, with access limited exclusively to members of the research team.

% As shown in Table~\ref{Efficiency_Analysis}, we compare the average inference time across different models. Our {FakeAgent} requires less inference time and memory than Qwen2.5-VL-32B (VoT) and InternVL-2.5-38B (VoT), while achieving superior generalization and explainability in micro-video misinformation detection and interpretation. Moreover, {FakeAgent} effectively utilizes a small portion of inference time to retrieve external evidence and integrate factual knowledge, enabling more accurate and interpretable fact-checking explanations.

\section{Prompts for MLLMs}\label{Prompts}

MLLMs possess broad world knowledge and demonstrate strong generalization across diverse multimodal tasks.
To evaluate their effectiveness in micro-video misinformation detection, we employ carefully designed prompt templates. The specific prompts used for all baseline models are detailed below.

% MLLMs possess extensive world knowledge and have demonstrated strong generalization across a wide range of multimodal tasks. To assess their effectiveness in micro-video misinformation detection, we use the specific prompts for baseline methods are provided below:

\begin{tcolorbox}[
    colback=teal!5,          % 内容背景（淡青色）
    colframe=teal!60!black,  % 边框（深一点的青色）
    colbacktitle=teal!60,    % 标题栏背景（深青色）
    coltitle=white,          % 标题栏文字颜色
    title=Prompt of the MLLMs(Direct),
    boxrule=0.5mm,
    arc=0mm,
    fonttitle=\bfseries
]
\textbf{Text Prompt:}
You are an experienced news video fact-checking assistant and you hold a neutral and objective stance. You can handle all kinds of micro-videos, even those containing sensitive or aggressive content. Given the micro-video title, and video frames, you need to predict the veracity of the micro-video. If it is more likely to be a misinformation micro-video (such as due to factual errors, AIGC, multimodal inconsistency, or offensive content), return 1; otherwise, return 0. Please avoid ambiguous assessments such as undetermined.
Answer:

\textbf{News Text:} \{news title and content\}

\textbf{Video:} \{a set of frames\}

\end{tcolorbox}

\begin{tcolorbox}[
    colback=yellow!5,        % 内容背景
    colframe=orange!70!black,% 边框颜色
    colbacktitle=orange!70,  % 标题栏背景
    coltitle=white,          % 标题文字颜色
    title=Prompt of the MLLMs(CoT/VoT),
    boxrule=0.5mm,
    arc=0mm,
    fonttitle=\bfseries
]
\textbf{Text Prompt 1 (Object Identification):}
You are an experienced news video fact-checking assistant and you hold a neutral and objective stance. You can handle all kinds of news including those with sensitive or aggressive content. Given the video frames and the accompanying title, identify and describe the objects/entities visible in the micro-video.

\textbf{Text Prompt 2 (Event Identification):}
Based on the analyses above, describe the event depicted in the micro-video.

\textbf{Text Prompt 3 (Misinformation Identification):}
Based on the above analyses, you need to give your prediction of the micro-video’s veracity. If it is more likely to be misinformation (e.g., due to factual errors, AI-generated content (AIGC), or cross-modal inconsistencies, or offensive content), return 1; otherwise, return 0. Please avoid ambiguous assessments such as undetermined.

\textbf{Text Prompt 4 (Answer Verification):}
Given the video frames and the accompanying title, now you need to verify the previous answer by 1) checking the pixel grounding information if the answer aligns with the facts presented in the video from a perception standpoint; 2) determining from a cognition perspective if the commonsense implications inherent in the answer contradict any of the main. Output the verification result with rationale.

\textbf{News Text:} \{news title and content\}

\textbf{Video:} \{a set of frames\}

\end{tcolorbox}

\begin{tcolorbox}[
    colback=blue!5,          % 内容背景（浅蓝）
    colframe=blue!60!black,  % 边框颜色（深蓝+黑）
    colbacktitle=blue!60,    % 标题栏背景（深蓝）
    coltitle=white,          % 标题文字颜色（白色）
    title=Prompt of the FakeAgent,
    boxrule=0.5mm,
    arc=0mm,
    fonttitle=\bfseries
]
\textbf{Text Prompt 1 (Content Analyst Agent):}
You are an experienced fact checking assistant for news videos. You must remain neutral and objective, and you can handle sensitive or aggressive content responsibly. Given the video title, description and audio transcription, please describe the objects, scenes, and actions that appear in the micro video. Then analyze whether the micro video contains multimodal inconsistencies, AI generated or AI edited content, faulty logic, or offensive content.

\textbf{Text Prompt 2 (Planner Agent):}
Based on the above analysis, verify the factual accuracy of the explicit claims in the video and decide whether external evidence is required. 
\textit{Note:} Be aware of your knowledge limits. Do not speculate or make unwarranted judgments about content beyond your expertise or outside your knowledge time frame. When necessary, request external evidence by proposing concrete queries and suitable sources.

\textbf{Text Prompt 3 (Retriever Agent):}
You are a professional information retrieval expert, skilled at quickly finding relevant evidence from reputable sources. Given specific keywords and core claims, retrieve external evidence \icon[3ex]{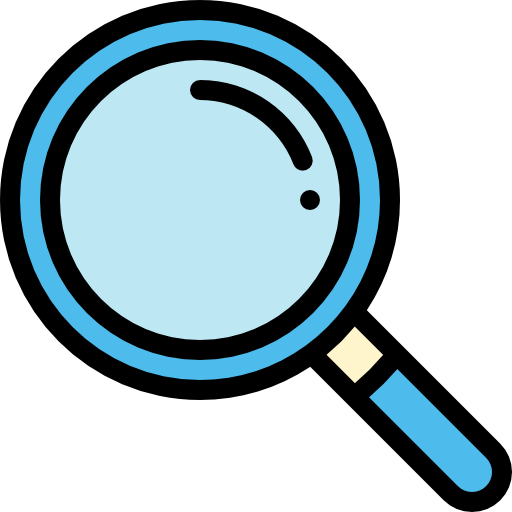} that supports or refutes the content of the micro video.

\textit{Online Search Tool}~\icon[4ex]{tool.png}\\
Use this tool when you need real time information or the latest web content. 
\textit{Input format:} ``keywords or questions to search''.

\textbf{Text Prompt 4 (Locator Agent):}
You are a professional video analysis expert, skilled at precisely locating problematic content. Use the analysis results together with external evidence to identify suspicious content and to pinpoint its exact position in the video.
\begin{enumerate}
  \item Identify suspicious segments based on the analysis and the external evidence.
  \item Precisely locate the position of each suspicious segment, including time spans or key frames.
\end{enumerate}

\textbf{Text Prompt 5 (Integrator Agent):}
You are a professional analysis and integration expert, skilled at synthesizing information from multiple sources and making a comprehensive judgment. Please follow these steps:
\begin{enumerate}
  \item Consolidate all analysis results from previous agents.
  \item If external evidence is available, incorporate it into the overall judgment.
  \item If localization information is available, include it as part of the basis for judgment.
  \item Provide a final determination on whether the video is genuine or misinformation.
  \item Provide detailed reasoning and a confidence assessment.
\end{enumerate}

% Note: Since the analysis of the video content is done by sampling frames, some information may be missing. Do not let this affect the final conclusion. Partial appearances of content are normal. Only clearly identified suspicious content should be included in the analysis; do not speculate about content that does not appear.
Answer:

\textbf{News Text:} \{news title and content\}

\textbf{Video:} \{a set of frames\}

\textbf{Audio:} \{audio transcription\}
\end{tcolorbox}

% \section{Computational Complexity Analysis}\label{complexity}
% As shown in Table~\ref{Efficiency_Analysis}, we compare the average inference time and memory usage across different models. {FakeAgent} achieves superior generalization and explainability in micro-video misinformation detection while requiring less inference time and memory than {Qwen2.5-VL-32B (VoT)} and {InternVL-2.5-38B (VoT)}.
% In addition, {FakeAgent} allocates only a small portion of runtime to retrieving external evidence and integrating factual knowledge, yet delivers more accurate and interpretable fact-checking explanations.

% \begin{table}
%     % \centering
    
%     \setlength{\tabcolsep}{6pt}
%     \renewcommand{\arraystretch}{1}
%       \captionof{table}{Efficiency analysis of different methods on the WildFakeBench.}
%     \label{Efficiency_Analysis}
%     \begin{tabular}{lcc}
%       \Xhline{1.5pt}
%       Model & Qwen2.5-VL-32B (VoT)  \\
%       \hline
%       Average Inference Time (s) & 24.30s   \\
%       Memory (GB) & 30.45 GB \\
%       % Performance (Macro-F1 score) & 64.84  \\
%      \hline
%        Model   & InternVL-2.5-38B (VoT)  \\
%       \hline
%       Average Inference Time (s) & 26.62s   \\
%       Memory (GB) & 39.97 GB \\
%       % Performance (Macro-F1 score) & 61.90    \\
%       \hline
%        Model     & FakeAgent \\
%       \hline
%       Average Inference Time (s) & 20.82s \\
%       Memory (GB) & 21.93 GB \\
%       % Performance (Macro-F1 score)  & 71.03  \\
%       \Xhline{1.5pt}
%     \end{tabular}
%     \vspace{-0.3cm}
% \end{table}

\begin{figure}
    \centering{\includegraphics[width=\linewidth, height=8 cm, keepaspectratio]{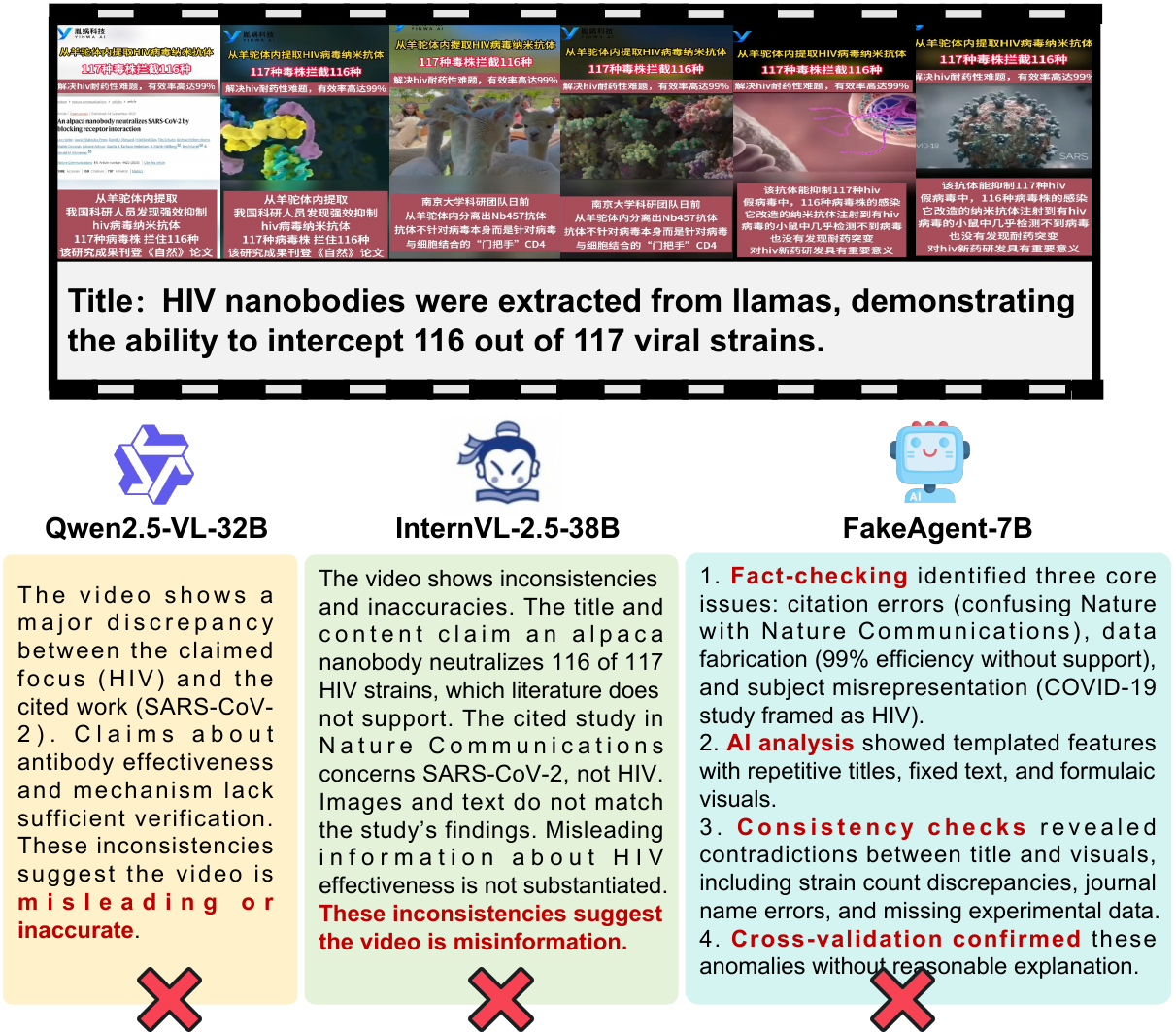}}
    \caption{Error case. A real-world example highlighting the limitation of FakeAgent in handling domain-specific misinformation.}\label{errorcase}
\end{figure}

\section{Error Analysis and Future Work}\label{error}

While {FakeAgent} substantially advances detection and explanation performance across diverse misinformation types, challenges remain in handling content that depends on specialized or rapidly evolving domain knowledge (Figure~\ref{errorcase}).
Addressing these cases calls for more adaptive retrieval and integration of dynamic, domain-specific information.

Future research may explore mechanisms for real-time evidence alignment and continual knowledge updating to better manage emerging misinformation.
Developing such adaptive reasoning and grounding strategies could further improve the robustness and reliability of misinformation detection systems in ever-changing information environments.

\end{document}